\documentclass[aps,prl,preprint,tightenlines,superscriptaddress,showpacs,byrevtex]{revtex4}

\usepackage{delarray}
\usepackage{multirow}
\usepackage{amsmath,amssymb}
\usepackage[dvipdfm]{graphicx}
\usepackage{wrapfig}
\usepackage{color}
\usepackage{subfig}
\usepackage{graphicx} 
\usepackage{dcolumn}  
\usepackage{lscape}
\usepackage{setspace}
\usepackage{textcomp}
\usepackage{feynmp}
\usepackage{here}
\graphicspath{{ps}}

\allowdisplaybreaks[3]

\newcommand{\mydif}{\mathrm{d}}
\newcommand{\bvec}[1]{\mbox{\boldmath $#1$}}

\usepackage[pdftitle={Measurement of etabar and xikappa of tau},pdfauthor={Nobuhiro Shimizu}]{hyperref}

\begin{document}

\preprint{\vbox{ \hbox{   }
                 \hbox{BELLE-CONF-1610}
                 \hbox{For the Tau 2016 conference}
}}

\title{ \quad\\[0.5cm]  Measurement of Michel Parameters ($\bar\eta$, $\xi\kappa$)\\ in the radiative leptonic decay $\tau \rightarrow l \nu \bar{\nu} \gamma$ at Belle}

\noaffiliation
\affiliation{Aligarh Muslim University, Aligarh 202002}
\affiliation{University of the Basque Country UPV/EHU, 48080 Bilbao}
\affiliation{Beihang University, Beijing 100191}
\affiliation{University of Bonn, 53115 Bonn}
\affiliation{Budker Institute of Nuclear Physics SB RAS, Novosibirsk 630090}
\affiliation{Faculty of Mathematics and Physics, Charles University, 121 16 Prague}
\affiliation{Chiba University, Chiba 263-8522}
\affiliation{Chonnam National University, Kwangju 660-701}
\affiliation{University of Cincinnati, Cincinnati, Ohio 45221}
\affiliation{Deutsches Elektronen--Synchrotron, 22607 Hamburg}
\affiliation{University of Florida, Gainesville, Florida 32611}
\affiliation{Department of Physics, Fu Jen Catholic University, Taipei 24205}
\affiliation{Justus-Liebig-Universit\"at Gie\ss{}en, 35392 Gie\ss{}en}
\affiliation{Gifu University, Gifu 501-1193}
\affiliation{II. Physikalisches Institut, Georg-August-Universit\"at G\"ottingen, 37073 G\"ottingen}
\affiliation{SOKENDAI (The Graduate University for Advanced Studies), Hayama 240-0193}
\affiliation{Gyeongsang National University, Chinju 660-701}
\affiliation{Hanyang University, Seoul 133-791}
\affiliation{University of Hawaii, Honolulu, Hawaii 96822}
\affiliation{High Energy Accelerator Research Organization (KEK), Tsukuba 305-0801}
\affiliation{J-PARC Branch, KEK Theory Center, High Energy Accelerator Research Organization (KEK), Tsukuba 305-0801}
\affiliation{Hiroshima Institute of Technology, Hiroshima 731-5193}
\affiliation{IKERBASQUE, Basque Foundation for Science, 48013 Bilbao}
\affiliation{University of Illinois at Urbana-Champaign, Urbana, Illinois 61801}
\affiliation{Indian Institute of Science Education and Research Mohali, SAS Nagar, 140306}
\affiliation{Indian Institute of Technology Bhubaneswar, Satya Nagar 751007}
\affiliation{Indian Institute of Technology Guwahati, Assam 781039}
\affiliation{Indian Institute of Technology Madras, Chennai 600036}
\affiliation{Indiana University, Bloomington, Indiana 47408}
\affiliation{Institute of High Energy Physics, Chinese Academy of Sciences, Beijing 100049}
\affiliation{Institute of High Energy Physics, Vienna 1050}
\affiliation{Institute for High Energy Physics, Protvino 142281}
\affiliation{Institute of Mathematical Sciences, Chennai 600113}
\affiliation{INFN - Sezione di Torino, 10125 Torino}
\affiliation{Advanced Science Research Center, Japan Atomic Energy Agency, Naka 319-1195}
\affiliation{J. Stefan Institute, 1000 Ljubljana}
\affiliation{Kanagawa University, Yokohama 221-8686}
\affiliation{Institut f\"ur Experimentelle Kernphysik, Karlsruher Institut f\"ur Technologie, 76131 Karlsruhe}
\affiliation{Kavli Institute for the Physics and Mathematics of the Universe (WPI), University of Tokyo, Kashiwa 277-8583}
\affiliation{Kennesaw State University, Kennesaw, Georgia 30144}
\affiliation{King Abdulaziz City for Science and Technology, Riyadh 11442}
\affiliation{Department of Physics, Faculty of Science, King Abdulaziz University, Jeddah 21589}
\affiliation{Korea Institute of Science and Technology Information, Daejeon 305-806}
\affiliation{Korea University, Seoul 136-713}
\affiliation{Kyoto University, Kyoto 606-8502}
\affiliation{Kyungpook National University, Daegu 702-701}
\affiliation{\'Ecole Polytechnique F\'ed\'erale de Lausanne (EPFL), Lausanne 1015}
\affiliation{P.N. Lebedev Physical Institute of the Russian Academy of Sciences, Moscow 119991}
\affiliation{Faculty of Mathematics and Physics, University of Ljubljana, 1000 Ljubljana}
\affiliation{Ludwig Maximilians University, 80539 Munich}
\affiliation{Luther College, Decorah, Iowa 52101}
\affiliation{University of Maribor, 2000 Maribor}
\affiliation{Max-Planck-Institut f\"ur Physik, 80805 M\"unchen}
\affiliation{School of Physics, University of Melbourne, Victoria 3010}
\affiliation{Middle East Technical University, 06531 Ankara}
\affiliation{University of Miyazaki, Miyazaki 889-2192}
\affiliation{Moscow Physical Engineering Institute, Moscow 115409}
\affiliation{Moscow Institute of Physics and Technology, Moscow Region 141700}
\affiliation{Graduate School of Science, Nagoya University, Nagoya 464-8602}
\affiliation{Kobayashi-Maskawa Institute, Nagoya University, Nagoya 464-8602}
\affiliation{Nara University of Education, Nara 630-8528}
\affiliation{Nara Women's University, Nara 630-8506}
\affiliation{National Central University, Chung-li 32054}
\affiliation{National United University, Miao Li 36003}
\affiliation{Department of Physics, National Taiwan University, Taipei 10617}
\affiliation{H. Niewodniczanski Institute of Nuclear Physics, Krakow 31-342}
\affiliation{Nippon Dental University, Niigata 951-8580}
\affiliation{Niigata University, Niigata 950-2181}
\affiliation{University of Nova Gorica, 5000 Nova Gorica}
\affiliation{Novosibirsk State University, Novosibirsk 630090}
\affiliation{Osaka City University, Osaka 558-8585}
\affiliation{Osaka University, Osaka 565-0871}
\affiliation{Pacific Northwest National Laboratory, Richland, Washington 99352}
\affiliation{Panjab University, Chandigarh 160014}
\affiliation{Peking University, Beijing 100871}
\affiliation{University of Pittsburgh, Pittsburgh, Pennsylvania 15260}
\affiliation{Punjab Agricultural University, Ludhiana 141004}
\affiliation{Research Center for Electron Photon Science, Tohoku University, Sendai 980-8578}
\affiliation{Research Center for Nuclear Physics, Osaka University, Osaka 567-0047}
\affiliation{Theoretical Research Division, Nishina Center, RIKEN, Saitama 351-0198}
\affiliation{RIKEN BNL Research Center, Upton, New York 11973}
\affiliation{Saga University, Saga 840-8502}
\affiliation{University of Science and Technology of China, Hefei 230026}
\affiliation{Seoul National University, Seoul 151-742}
\affiliation{Shinshu University, Nagano 390-8621}
\affiliation{Showa Pharmaceutical University, Tokyo 194-8543}
\affiliation{Soongsil University, Seoul 156-743}
\affiliation{University of South Carolina, Columbia, South Carolina 29208}
\affiliation{Stefan Meyer Institute for Subatomic Physics, Vienna 1090}
\affiliation{Sungkyunkwan University, Suwon 440-746}
\affiliation{School of Physics, University of Sydney, New South Wales 2006}
\affiliation{Department of Physics, Faculty of Science, University of Tabuk, Tabuk 71451}
\affiliation{Tata Institute of Fundamental Research, Mumbai 400005}
\affiliation{Excellence Cluster Universe, Technische Universit\"at M\"unchen, 85748 Garching}
\affiliation{Department of Physics, Technische Universit\"at M\"unchen, 85748 Garching}
\affiliation{Toho University, Funabashi 274-8510}
\affiliation{Tohoku Gakuin University, Tagajo 985-8537}
\affiliation{Department of Physics, Tohoku University, Sendai 980-8578}
\affiliation{Earthquake Research Institute, University of Tokyo, Tokyo 113-0032}
\affiliation{Department of Physics, University of Tokyo, Tokyo 113-0033}
\affiliation{Tokyo Institute of Technology, Tokyo 152-8550}
\affiliation{Tokyo Metropolitan University, Tokyo 192-0397}
\affiliation{Tokyo University of Agriculture and Technology, Tokyo 184-8588}
\affiliation{University of Torino, 10124 Torino}
\affiliation{Toyama National College of Maritime Technology, Toyama 933-0293}
\affiliation{Utkal University, Bhubaneswar 751004}
\affiliation{Virginia Polytechnic Institute and State University, Blacksburg, Virginia 24061}
\affiliation{Wayne State University, Detroit, Michigan 48202}
\affiliation{Yamagata University, Yamagata 990-8560}
\affiliation{Yonsei University, Seoul 120-749}
  \author{A.~Abdesselam}\affiliation{Department of Physics, Faculty of Science, University of Tabuk, Tabuk 71451} 
  \author{I.~Adachi}\affiliation{High Energy Accelerator Research Organization (KEK), Tsukuba 305-0801}\affiliation{SOKENDAI (The Graduate University for Advanced Studies), Hayama 240-0193} 
  \author{K.~Adamczyk}\affiliation{H. Niewodniczanski Institute of Nuclear Physics, Krakow 31-342} 
  \author{H.~Aihara}\affiliation{Department of Physics, University of Tokyo, Tokyo 113-0033} 
  \author{S.~Al~Said}\affiliation{Department of Physics, Faculty of Science, University of Tabuk, Tabuk 71451}\affiliation{Department of Physics, Faculty of Science, King Abdulaziz University, Jeddah 21589} 
  \author{K.~Arinstein}\affiliation{Budker Institute of Nuclear Physics SB RAS, Novosibirsk 630090}\affiliation{Novosibirsk State University, Novosibirsk 630090} 
  \author{Y.~Arita}\affiliation{Graduate School of Science, Nagoya University, Nagoya 464-8602} 
  \author{D.~M.~Asner}\affiliation{Pacific Northwest National Laboratory, Richland, Washington 99352} 
  \author{T.~Aso}\affiliation{Toyama National College of Maritime Technology, Toyama 933-0293} 
  \author{H.~Atmacan}\affiliation{Middle East Technical University, 06531 Ankara} 
% ...
\collaboration{The Belle Collaboration}

\begin{abstract}
We present the first measurement of the Michel parameters $\bar{\eta}$ and $\xi\kappa$ in the radiative leptonic
 decay of the $\tau$ lepton using 703~f$\mathrm{b}^{-1}$ of data collected with the Belle detector at the KEKB $e^+e^-$ collider.
 The Michel parameters are measured by an unbinned maximum likelihood fit to the kinematic information
 of $e^+e^-\rightarrow\tau^+\tau^-\rightarrow (\pi^+\pi^0 \bar{\nu})(l^-\nu\bar{\nu}\gamma)$
 $(l=e$ or $\mu)$. The preliminary values of the measured Michel parameters are $\bar{\eta} = -2.0 \pm 1.5 \pm 0.8$ and $\xi\kappa = 0.6 \pm 0.4 \pm 0.2$,
 where the first error is statistical and the second
 is systematic.
\end{abstract}

\maketitle

\tighten

\newpage
\section{Introduction}

In the Standard Model (SM), there are three flavors of charged leptons: $e, \mu$ and $\tau$.
 The SM has proven to be the most powerful theory in describing the physics of leptons; nevertheless, precision tests of the SM may reveal the presence of New Physics (NP).
 In particular, measurement of the Michel parameters in $\tau$ decay is a powerful probe for NP.  

The most general Lorentz-invariant derivative-free matrix element of leptonic $\tau$ decay
 $\tau^-\rightarrow l^- \nu \bar{\nu} \gamma$~\footnotemark[1] \footnotetext[1]{Unless otherwise stated, use of charge-conjugate modes is implied
throughout the paper.} is represented as~\cite{MPform} \\ \\

\hspace{2mm}  \raisebox{1cm}{$\mathcal{M} = $} 
\begin{fmffile}{tautree}
\begin{fmfchar*}(90,60)
  \fmfleft{tm,antinu} \fmflabel{$\tau$}{tm} \fmflabel{$\nu_{\tau}$}{antinu}
  \fmf{fermion}{tm,Wi}
  \fmf{fermion}{Wi,antinu}
  \fmf{photon}{Wi,Wf}
  \fmf{fermion,label=\rotatebox{60}{\vspace{-2mm}\hspace{-3mm}$~$}}{fb,Wf}
  \fmf{fermion}{Wf,f}
  \fmfright{f,fb} \fmflabel{$\nu_{l}$}{fb} \fmflabel{$l$}{f}
  \fmfdot{Wi,Wf}
\end{fmfchar*}
\end{fmffile}
\begin{equation}
\hspace{-4cm}\raisebox{-0.5cm}{$=\displaystyle \frac{4G_F}{\sqrt{2}}\sum_{\substack{N=S,V,T \\ i,j=L,R}}g_{ij}^{N}\left[\overline{u}_{i}(l)\Gamma^{N}v_{n}(\nu_l)\right]\left[\overline{u}_{m}(\nu_{\tau})\Gamma_{N}u_{j}(\tau)\right],$}
\end{equation}
where $G_F$ is the Fermi constant, $i$ and $j$ are the chirality
indices for the charged leptons, $n$ and $m$ are the chirality indices
of the neutrinos, $l$ is $e$ or $\mu$, $\Gamma^{S}=1$,
$\Gamma^{V}=\gamma^{\mu}$ and $\displaystyle
\Gamma^{T}={i}\left(\gamma^{\mu}\gamma^{\nu}-\gamma^{\nu}\gamma^{\mu}\right)/2\sqrt{2}$
are, respectively, the scalar, vector and tensor Lorentz structures
in terms of the Dirac matrices $\gamma^{\mu}$, $u_i$ and $v_i$ are the four-component spinors of
 a particle and an antiparticle, respectively and $g_{ij}^{N}$ are the
corresponding dimensionless couplings. In the SM, $\tau^-$ decays into $l^-$
 via the $W^-$ vector boson with a right-handed antineutrino, \textit{i.e.,} the only non-zero coupling is $g_{LL}^{V}=1$.
 Experimentally, only the squared matrix element is observable and so bilinear combinations of the $g_{ij}^{N}$ are accessible.
Of all such combinations, four Michel parameters, $\eta $, $\rho $, $\delta $ and $\xi $, can be measured by the leptonic decay of the $\tau$
 when the final state neutrinos are not observed~\cite{ordMP}:
\begin{eqnarray}
\displaystyle \rho&=&\frac{3}{4}-\frac{3}{4}\left(\left|g_{LR}^{V}\right|^{2}+\left|g_{RL}^{V}\right|^{2} +2\left|g_{LR}^{T}\right|^{2}+2\left|g_{RL}^{T}\right|^{2} +    \Re\left(g_{LR}^{S}g_{LR}^{T*}+g_{RL}^{S}g_{RL}^{T*} \right)\right) ,\\
\displaystyle \eta&=&\frac{1}{2}\Re\left(6g_{RL}^{V}g_{LR}^{T*}+6g_{LR}^{V}g_{RL}^{T*}+g_{RR}^{S}g_{LL}^{V*}+g_{RL}^{S}g_{LR}^{V*}+g_{LR}^{S}g_{RL}^{V*}+g_{LL}^{S}g_{RR}^{V*}\right) ,\\
\displaystyle \xi&=&4\Re\left(g_{LR}^{S}g_{LR}^{T*}-g_{RL}^{S}g_{RL}^{T*}\right)+\left|g_{LL}^{V}\right|^{2}+3\left|g_{LR}^{V}\right|^{2}-3\left|g_{RL}^{V}\right|^{2}-\left|g_{RR}^{V}\right|^{2} \nonumber \\
&{}&+5\left|g_{LR}^{T}\right|^{2}-5\left|g_{RL}^{T}\right|^{2}+\frac{1}{4}\left(\left|g_{LL}^{S}\right|^{2}-\left|g_{LR}^{S}\right|^{2}+\left|g_{RL}^{S}\right|^{2}-\left|g_{RR}^{S}\right|^{2}\right) ,\\
\displaystyle \xi\delta&=&\frac{3}{16}\left(\left|g_{LL}^{S}\right|^{2}-\left|g_{LR}^{S}\right|^{2}+\left|g_{RL}^{S}\right|^{2}-\left|g_{RR}^{S}\right|^{2}\right) \nonumber \\
&{}&-\frac{3}{4}\left(\left|g_{LR}^{T}\right|^{2}-\left|g_{RL}^{T}\right|^{2}-\left|g_{LL}^{V}\right|^{2}+\left|g_{RR}^{V}\right|^{2}-\Re\left(g_{LR}^{S}g_{LR}^{T*}+g_{RL}^{S}g_{RL}^{T*}\right)\right).
\end{eqnarray}
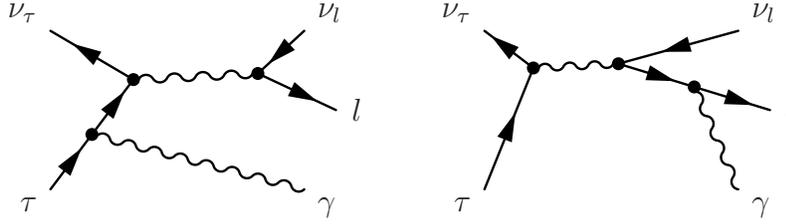
\begin{figure}[]
\begin{center}
\begin{fmffile}{externalph1}
\begin{fmfchar*}(120,60)
  \fmfleft{tm3,antinu3} \fmflabel{$\tau$}{tm3} \fmflabel{$\nu_{\tau}$}{antinu3}
  \fmfright{g3,f3,fb3} \fmflabel{$\nu_{l}$}{fb3} \fmflabel{$l$}{f3} \fmflabel{$\gamma $}{g3}
  \fmf{fermion}{tm3,v23,Wi3,antinu3}
  \fmf{photon,tension=1}{Wi3,Wf3}
  \fmf{photon,tension=0}{v23,g3}
  \fmf{fermion,label=\rotatebox{50}{\vspace{-2mm}\hspace{-3mm}$~$}}{fb3,Wf3}
  \fmf{fermion}{Wf3,f3}
  \fmfdot{Wi3,Wf3,v23}
\end{fmfchar*}
\end{fmffile}\hspace{10mm}
\begin{fmffile}{externalph2}
\begin{fmfchar*}(120,60)
  \fmfleft{tm4,antinu4} \fmflabel{$\tau$}{tm4} \fmflabel{$\nu_{\tau}$}{antinu4}
  \fmfright{g4,f4,fb4} \fmflabel{$\nu_{l}$}{fb4} \fmflabel{$l$}{f4} \fmflabel{$\gamma $}{g4}
  \fmf{fermion}{tm4,Wi4}
  \fmf{fermion,tension=3}{Wi4,antinu4}
  \fmf{photon,tension=2.3}{Wi4,Wf4}
  \fmf{fermion,label=\rotatebox{13}{\vspace{-2mm}\hspace{-3mm}$~$}}{fb4,Wf4}
  \fmf{fermion}{Wf4,v24,f4}
  \fmf{photon,tension=0}{v24,g4}
  \fmfdot{Wi4,Wf4,v24}
\end{fmfchar*}
\end{fmffile}
\end{center}
\caption{Feynman diagram of the radiative leptonic decay of the $\tau$ lepton.}\label{RD}
\end{figure}
The Feynman diagrams describing the radiative leptonic decay of the
$\tau$ are presented in Fig.~\ref{RD}. As shown in Refs.~\cite{oldform,michel_form_arvzov},
the presence of a radiative photon in the final state ({\it radiative
leptonic decay} or {\it inner bremsstrahlung}) exposes three more Michel
parameters, $\bar{\eta}$, $\eta^{\prime \prime}$ and $\xi\kappa$:
\begin{eqnarray}
\bar{\eta}&=&\left|g_{RL}^{V}\right|^{2}+\left|g_{LR}^{V}\right|^{2}+\frac{1}{8}\left(\left|g_{RL}^{S}+2g_{RL}^{T}\right|^{2}+\left|g_{LR}^{S}+2g_{LR}^{T}\right|^{2}\right)+2\left(\left|g_{RL}^{T}\right|^{2}+\left|g_{LR}^{T}\right|^{2}\right) \label{etaform}, \\
\eta^{\prime\prime}&=&\Re\left\{ 24g_{RL}^{V}(g_{LR}^{S*}+6g_{LR}^{T*})+24g_{LR}^{V}(g_{RL}^{S*}+6g_{RL}^{T*})-8(g_{RR}^{V}g_{LL}^{S*}+g_{LL}^{V}g_{RR}^{S*})\right\}, \\
\xi\kappa&=&\left|g_{RL}^{V}\right|^{2}-\left|g_{LR}^{V}\right|^{2}+\frac{1}{8}\left(\left|g_{RL}^{S}+2g_{RL}^{T}\right|^{2}-\left|g_{LR}^{S}+2g_{LR}^{T}\right|^{2}\right)+2\left(\left|g_{RL}^{T}\right|^{2}-\left|g_{LR}^{T}\right|^{2}\right).
\end{eqnarray}
Both $\bar{\eta}$ and $\eta^{\prime \prime}$ appear as spin-independent terms in the differential decay width.
 Since all terms in Eq.~\ref{etaform} are strictly non-negative, the upper limit on $\bar{\eta}$ provides a constraint on each coupling constant.
 The effect of the nonzero value of $\eta^{\prime \prime}$ is suppressed by a factor of $m_l^2/m_\tau^2 \sim 10^{-7}$ for an electron
 daughter and $\sim 0.4\%$ for a muon daughter
 and so proves to be difficult to measure with the available statistics of the Belle experiment. In this study, we use the SM value $\eta^{\prime \prime} =0$.
 
To measure $\xi\kappa$, which appears in the spin-dependent part of the differential decay width, we must determine the spin direction of the $\tau$.
 This spin dependence is extracted using the spin-spin correlation with the partner $\tau$ in the event.
 According to Ref.~\cite{xiprelation},
 $\xi\kappa$ is related to another Michel-like parameter $\xi^\prime = -\xi -4\xi\kappa + 8\xi\delta/3$.
 Because the normalized probability that the $\tau^-$ decays into the right-handed charged daughter lepton $Q_R^\tau$
 is given by $Q_R^\tau=(1-\xi^\prime)/2$~\cite{xiprimeapll}, the measurement of $\xi\kappa$ provides
 a further constraint on the $V-A$ structure of the weak current. 
The information on these parameters is summarized in Table~\ref{MPs}.

 Using the statistically abundant data set of ordinary leptonic decays, previous measurements~\cite{etalep,rhoCLEO} had determined the Michel
 parameters $\eta$, $\rho$, $\delta$ and $\xi$ to an accuracy of a few percent and in agreement with the SM prediction.
 Taking into account this measured agreement, the smaller data set of the radiative decay and its limited sensitivity, we focus in this analysis
 only on the extraction of $\bar{\eta }$ and $\xi\kappa$ by fixing $\eta$, $\rho$, $\delta$ and $\xi$ to the SM values.
 This represents the first measurement of the $\bar{\eta }$ and $\xi\kappa$ parameters of the $\tau$ lepton.

\begin{table}[]
\caption{Michel parameters of the $\tau$ lepton}\label{MPs}
\begin{center}
\begin{tabular}{ccccc} \hline \hline
Name &SM&Spin & Experimental& Comments and Ref. \\  
          &value& correlation &    result  $^\dagger$ \cite{PDG_paper}    &                  \\ \hline 
$\eta$& 0 & no & $0.013\pm0.020$ & \cite{etalep} \\ 
$\rho $& $3/4$ & no & $0.745 \pm 0.008$ & \cite{rhoCLEO} \\
$\xi\delta $& $3/4$ & yes &$0.746\pm 0.021$ &\cite{rhoCLEO}\\
$\xi $& 1 & yes & $0.995\pm0.007$ & measured in hadronic decays \cite{etalep}\\
$\overline{\eta }$& 0 & no & not measured & ~~from radiative decay (RD)\\
$\xi\kappa$& 0 & yes & not measured& from RD\\  
$\eta^{\prime \prime}$& 0 & no & not measured & ~~~~~~~from RD, suppressed by $m_l^2/m_\tau^2$\\
$\xi^{\prime}$& 1 & yes & - & induced from $\xi^\prime = -\xi -4\xi\kappa + 8\xi\delta/3$. \\  \hline \hline
\end{tabular}
\end{center}
\begin{flushleft}\vspace{-3mm}$~^\dagger$~{\small Experimental results represent average values obtained by PDG~\cite{PDG_paper}. The most precise results are referenced here.}\end{flushleft}
\end{table}

\section{Method}
Hereafter, we use an italic character to represent the four-vector $p$ while its time and spatial components are denoted by capital letters as $p=(E, \bvec{P})$.
 The magnitude of $\bvec{P}$ is denoted as $P$.
 
The differential decay width for the radiative leptonic
 decay of $\tau^-$ with a definite spin direction $\bvec{S}_{\tau^-}$ is given by
\begin{equation}
\frac{\mydif\Gamma(\tau^-\rightarrow l^- \nu \bar{\nu} \gamma )}{\mydif E^*_l \mydif \Omega^*_l \mydif E^*_\gamma \mydif \Omega^*_\gamma } 
= \left( A_{0}^-+\bar{\eta}\,A_{1}^- \right)+ \left( \bvec{B}_{0}^- +\xi\kappa\,\bvec{B}_{1}^- \right)\cdot \bvec{S}_{\tau^-},\label{taum}
\end{equation}
where $A_{i}^{-}$ and $\bvec{B}_{i}^-$~~are known functions of
the kinematics of the decay products with indices $i=0,1$ ($i$ simply denotes the name of function),
 $\Omega_a$ stands for a set of $\{\mathrm{cos}\theta_{a}, \phi_{a}\}$ for a particle type $a$ and the asterisk means that the variable is defined in the
 $\tau$ rest frame. The detailed formula is given in Appendix~A. Equation~\ref{taum} shows that
$\xi\kappa$ appears in the {\it spin-dependent} part of the decay
width. This product can be measured by utilizing
the well-known spin-spin correlation of the $\tau$ pair in the
$e^-e^+\rightarrow\tau^+\tau^-$ reaction:
\begin{eqnarray}
 \frac{\mathrm{d} \sigma \left( e^-e^+ \rightarrow \tau^- (\bvec{S}^{-}) \tau^+ ( \bvec{S}^{+} ) \right) }{\mathrm{d}\Omega_{\tau}}
= \frac{\alpha^2\beta_{\tau}}{64E_{\tau}^2}(D_{0}+{\textstyle \sum_{i,j}} D_{ij}S^-_iS^+_j)
,\label{corr}
\end{eqnarray}
where $\alpha$ is the fine structure constant, $\beta_{\tau}$ and
$E_{\tau}$ are the velocity and energy of the $\tau$, respectively,
$D_{0}$ is a form factor for the spin-independent part of the reaction
and $D_{ij}$ $(i,j=0,1,2)$ is a tensor describing the spin-spin
correlation~\cite{Tsai}:
\begin{equation} \scalebox{1}{
$D_0 = 1+\cos^2{\theta}+\displaystyle \frac{1}{\gamma^2_{\tau}}\sin^2{\theta}$,
}\end{equation}
\begin{equation}
D_{ij} = \left( \begin{array}{@{}c@{~~}c@{~~}c@{}}
(1+{\displaystyle 1 \over \gamma_{\tau}^2})\sin^2{\theta} & 0 & {\displaystyle 1 \over \gamma_{\tau}}\sin{2\theta} \\
 0 & -\beta^2_{\tau}\sin^2{\theta} & 0 \\ 
{\displaystyle 1 \over \gamma_{\tau}}\sin{2\theta} & 0 & 1+\cos^2{\theta}-{\displaystyle 1 \over \gamma_{\tau}^2}\sin^2{\theta} \\ 
\end{array} \right);
\end{equation}
here, $\theta$ is the polar angle of the $\tau^-$ and $\gamma_{\tau}$ is its gamma factor $1/\sqrt{1-\beta_{\tau}^2}$. 

The spin information on the partner $\tau^+$ is extracted using the two-body decay $\tau^+\rightarrow \rho^+ \bar{\nu} \rightarrow\pi^+\pi^0\bar{\nu}$
 whose differential decay width is
\begin{equation}
\frac{\mathrm{d}\Gamma (\tau^+\rightarrow
  \pi^+\pi^0\bar{\nu})}{\mydif \Omega_{\rho}^*\mydif m^2 \mydif \widetilde{\Omega}_{\pi}} = A^+ +\bvec{B}^+\cdot \bvec{S}_{\tau^+};
\end{equation}
$A^+$ and $\bvec{B}^+$ are the form
factors for the spin-independent and spin-dependent parts,
respectively, while the tilde indicates the variables defined in the 
$\rho$ rest frame and $m$
 is an invariant mass of the two-body system of pions which is defined as $m^2=(p_\pi+p_{\pi^0})^2$.
 The formulae of $A^+$ and $\bvec{B}^+$ are given in Appendix~B.
 Thus, the total differential cross section of
$e^+e^-\rightarrow\tau^-\tau^+\rightarrow
(l^-\nu \bar{\nu}\gamma)(\pi^+\pi^0\bar{\nu})$
{(or, briefly, $(l^{-}\gamma, \pi^{+}\pi^0)$) can be written as:
\begin{equation}
\displaystyle \frac{\mathrm{d}\sigma(l^-\gamma,\pi^+\pi^0) }{ \mydif E^*_l \mydif \Omega^*_l \mydif E^*_\gamma \mydif \Omega^*_\gamma  \mydif \Omega_{\rho}^* \mydif m^2 \mydif \widetilde{\Omega}_{\pi} \mydif \Omega_\tau} \propto \frac{\beta_{\tau}}{E_{\tau}^2}
\left[ D_{0} \left(A_{0}^-\! +A_{1}^-\!\cdot\!\bar{\eta} \right) A^+ + {\textstyle \sum_{i,j}} D_{ij} \left(\bvec{B}_{0}^-\!+\! \bvec{B}_{1}^-\!\cdot\!\xi\kappa \right)_i\cdot\! \bvec{B}^+_j \right].\label{totform}
\end{equation}
To extract the visible differential cross section, we transform the differential variables into ones defined
 in the center-of-mass system (CMS) using the Jacobian $J$
 ($\mydif E^*_l \mydif \Omega^*_l \mydif E^*_\gamma \mydif \Omega^*_\gamma  \mydif \Omega_{\rho}^* \mydif \Omega_{\tau}$  $\rightarrow$ $\mydif \Phi \mydif P_l \mydif \Omega_l \mydif P_\gamma \mydif \Omega_\gamma \mydif P_\rho \mydif \Omega_\rho  $): 
\begin{align}
&\hspace{3em}J = J_1 J_2 J_3, \\
&\hspace{3em}J_1= \left|\frac{\partial(E_{l}^{*}, \Omega_{l}^*)}{\partial(P_l, \Omega_l)} \right| = \frac{P_{l}^2}{E_l { P}_{l}^*}, \\
&\hspace{3em}J_2= \left|\frac{\partial(E_{\gamma}^{*}, \Omega_{\gamma}^*)}{\partial(P_\gamma, \Omega_\gamma)} \right| = \frac{E_\gamma}{E_\gamma^*}, \\
&\hspace{3em}J_3 = \left| \frac{\partial (\Omega_{\rho}^*, \Omega_{\tau})}{\partial ( P_{\rho}, \Omega_{\rho}, \Phi  )}\right|
=  \frac{m_{\tau} P_{\rho} }{ E_{\rho}P_{\rho}^* P_{\tau} },
\end{align}
where the parameter $\Phi$ denotes the angle along the arc illustrated in Fig.~\ref{constraingedcone}.
 On the assumption that the neutrino is massless and the invariant mass of the
 neutrino pair is greater than or equal to zero, we obtain
\begin{align}
0 &= p_{\bar{\nu}}^2 = (p_{\tau}-p_{\rho})^2 = m_\tau^2 + m^2 - 2 E_{\tau} E_\rho + 2 P_\tau P_\rho \mathrm{cos}\theta_{\tau \rho}, \\
0 &\leq p_{{\nu}\bar{\nu}}^2 = (p_{\tau}-p_{l\gamma})^2 = m_\tau^2 + m_{l\gamma}^2 - 2 E_{\tau} E_{l\gamma} + 2 P_{\tau} P_{l \gamma} \mathrm{cos}\theta_{\tau (l\gamma)}. 
\end{align}\vspace{-10mm}\begin{center}$\Longleftrightarrow  $\end{center} \vspace{-9mm}
\begin{align}
\mathrm{cos}\theta_{\tau \rho} &= \frac{ 2 E_{\tau} E_\rho -m_\tau^2 - m^2 }{2 P_\tau P_\rho } , \\
\mathrm{cos}\theta_{\tau (l\gamma)} &\geq  \frac{2 E_{\tau} E_{l\gamma}-m_\tau^2 - m_{l\gamma}^2 }{2 P_{\tau} P_{l \gamma} }. 
\end{align}
In the back-to-back topology of the $\tau^+\tau^-$ pair, these two conditions constrain the $\tau^+$ direction to the arc,
 with the angle $\Phi$ defined along this arc.
The visible differential cross section is, therefore, obtained by integration over $\Phi$:
\begin{align}
\displaystyle \frac{\mathrm{d}\sigma(l^-\gamma,\pi^+\pi^0) }{ \mydif P_l \mydif \Omega_l \mydif P_\gamma \mydif \Omega_\gamma \mydif P_\rho \mydif \Omega_\rho \mydif m^2 \mydif \widetilde{\Omega}_\pi   } 
&= \int_{\Phi_1}^{\Phi_2} \hspace{-0.5em} \mydif \Phi
\frac{\mathrm{d}\sigma(l^-\gamma,\pi^+\pi^0) }{\mydif \Phi \mydif P_l \mydif \Omega_l \mydif P_\gamma \mydif \Omega_\gamma \mydif P_\rho \mydif \Omega_\rho   \mydif m^2 \mydif \widetilde{\Omega}_{\pi} } \\
&= \int_{\Phi_1}^{\Phi_2} \hspace{-0.5em} \mydif \Phi
\frac{\mathrm{d}\sigma(l^-\gamma,\pi^+\pi^0) }{\mydif E^*_l \mydif \Omega^*_l \mydif E^*_\gamma \mydif \Omega^*_\gamma  \mydif \Omega_{\rho}^* \mydif m^2 \mydif \widetilde{\Omega}_{\pi} \mydif \Omega_\tau} J\label{totformwithJ} \\
&\equiv S(\bvec{x}),
\end{align}
where $S(\bvec{x})$ is proportional to the probability density function (PDF) of the signal and $\bvec{x}$ denotes the set of twelve measured variables:
 $\bvec{x} = \{ P_l, \Omega_l, P_\gamma, \Omega_\gamma, P_\rho, \Omega_\rho, m^2, \widetilde{\Omega}_\pi \}$.
 
In general, the normalization of the PDF depends on the Michel parameters. Since $S(\bvec{x})$ is a linear combination of the Michel parameters
 $S(\bvec{x})=A_0(\bvec{x}) + A_1(\bvec{x}) \bar{\eta} + A_2(\bvec{x}) \xi\kappa$,
 the PDF is normalized according to
\begin{align}
\frac{S(\bvec{x})}{\int \hspace{-0.3em} \mydif \bvec{x} S(\bvec{x})} = \frac{A_0(\bvec{x}) + A_1(\bvec{x}) \bar{\eta} + A_2(\bvec{x}) \xi\kappa}
{\int \hspace{-0.3em} \mydif \bvec{x} \left( A_0(\bvec{x}) + A_1(\bvec{x}) \bar{\eta} + A_2(\bvec{x}) \xi\kappa \right) } \equiv \frac{A_0(\bvec{x}) + A_1(\bvec{x}) \bar{\eta} + A_2(\bvec{x}) \xi\kappa}{N_0 + N_1 \bar{\eta} + N_2 \xi\kappa}, \label{normformula}
\end{align}
 where $N_i$ ($i=0,1,2$) is a normalization coefficient defined by $N_i=\int \hspace{-0.1em} \mydif \bvec{x} A_i(\bvec{x})$.
 This integration is performed using the Monte Carlo (MC) method.
 Since MC events are distributed according to the SM distribution ($\bar{\eta}=\xi\kappa=0$), the denominator of Eq.~\ref{normformula} is
\begin{align}
\int \hspace{-0.3em} \mydif \bvec{x} \left( A_0(\bvec{x}) + A_1(\bvec{x}) \bar{\eta} + A_2(\bvec{x}) \xi\kappa \right) &= N_0\int \hspace{-0.3em} \mydif \bvec{x} \left(\frac{A_0(\bvec{x})}{N_0} \right)
 \cdot \frac{ A_0(\bvec{x}) + A_1(\bvec{x}) \bar{\eta} + A_2(\bvec{x}) \xi\kappa }{A_0(\bvec{x})} \\
 &=  \frac{N_0}{N_{\rm gen}} \sum_{i:{\rm gen}} \frac{ A_0(\bvec{x}^i) + A_1(\bvec{x}^i) \bar{\eta} + A_2(\bvec{x}^i) \xi\kappa }{A_0(\bvec{x}^i)} \\
 &= N_0 \left[ 1+ \left< \frac{A_1}{A_0} \right> \bar{\eta} + \left< \frac{A_2}{A_0} \right> \xi\kappa \right],
\end{align}
where $\bvec{x}^i$ represents a set of variables for $i^{\rm th}$ generated event out of total $N_{\rm gen}$ events and the bracket $<>$ means an average with respect to the SM distribution.
 We refer to $N_0$ and $\left< {A_i}/{A_0} \right>$ ($i=1,2$) as absolute and relative normalizations, respectively.
\begin{figure}[]
{\centering\includegraphics[width=5.5cm]{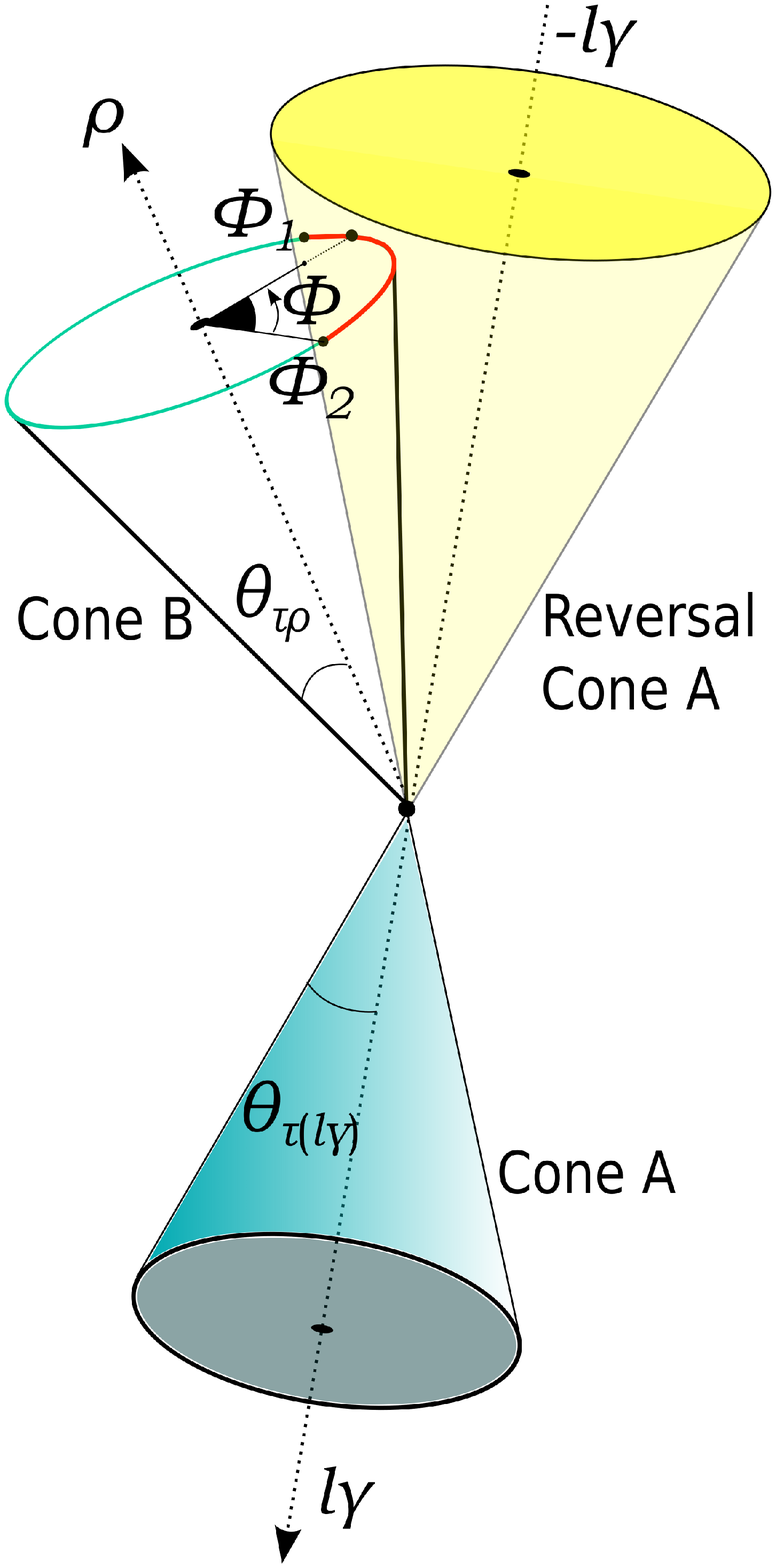}}
\caption[]{	\raggedright Kinematics of $\tau^+\tau^- \rightarrow (\rho^+\bar{\nu})(l^-\nu\bar{\nu}\gamma)$ decay.
 Cones A and B are the surfaces that satisfy the conditions $(p_{\tau^-}-p_{l^-\gamma})^2=0$ and $(p_{\tau^+}-p_{\rho^+})^2=0$ in the CMS frame.
 The direction of $\tau^+$ is constrained to lie on an arc defined by the intersection of the reversal (\textit{i.e.,} mirror) cone~A and cone~B.
 The arc (shown in red) is parametrized by an angle $\Phi\in[\Phi_1, \Phi_2]$. }\label{constraingedcone}
\end{figure}

\section{KEKB accelerator}
The KEKB accelerator, located at KEK in Tsukuba, Ibaraki, Japan, is an energy-asymmetric $e^+e^-$
 collider with beam energies of 3.5~GeV and 8.0~GeV for $e^+$ and $e^-$, respectively.
 Most of the data taking was in operation at the CMS energy of 10.58~GeV,
 a mass of the $\Upsilon (4S)$, where a huge number of $\tau^+\tau^-$
 as well as $B\overline{B}$ pairs were produced.
 The KEKB was operated from 1999 to 2010 and accumulated 1~${\rm ab^{-1}}$ of $e^+e^-$ collision data with Belle detector.
 The achieved instantaneous luminosity of $2.11\times 10^{34}{\rm~cm^{-2}/s}$ is the world-largest record.
 For this reason, the KEKB is often called $B$-{\it factory}
 but it is worth considering also as $\tau$-{\it factory}, where $O(10^9)$ events of $\tau$ pair have been produced.
 The large number of the $\tau$ leptons and the dedicated detector provide
 a beautiful laboratory for the test of the nature of the rare decay $\tau^- \rightarrow l^- \nu \bar{\nu} \gamma$.
 The KEKB is described in detail in Refs.~\cite{cite_KEKB1}.
\section{Belle detector}
The Belle detector is a large-solid-angle magnetic
spectrometer that consists of a silicon vertex detector (SVD),
a 50-layer central drift chamber (CDC), an array of
aerogel threshold Cherenkov counters (ACC),  
a barrel-like arrangement of time-of-flight
scintillation counters (TOF), and an electromagnetic calorimeter
comprised of CsI(Tl) crystals (ECL) located inside 
a superconducting solenoid coil that provides a 1.5~T
magnetic field.  An iron flux return located outside of
the coil is instrumented to detect $K_L^0$ mesons and to identify
muons (KLM). The detector
is described in detail elsewhere~\cite{citeBelle}.

\section{Event selection}

The selection proceeds in two stages. At the preselection, $\tau^+\tau^-$ candidates are selected efficiently while suppressing
 the beam background and other physics processes like Bhabha scattering, two-photon interaction and $\mu^+\mu^-$ pair production.
 The preselected events are then required to satisfy final selection criteria to enhance the purity of the radiative events.
 \vspace{-4mm}
\subsection{Preselection}
\begin{itemize}
\item {There must be exactly two oppositely charged tracks in the event.
 The impact parameters of these tracks relative to the interaction point are required to be
 within $\pm 2.5$~cm along the beam axis and $\pm0.5$~cm in the transverse plane.
 The transverse momentum of two tracks must exceed $0.1$~GeV/$c$ and that of one track must exceed $0.5$~GeV/$c$. }
\item{Total energy deposition of ECL in the laboratory frame must be lower than 9~GeV.}
\item{The opening angle $\psi$ of the two tracks must satisfy $20^\circ<\psi < 175^\circ$.}
\item{The number of photons whose energy exceeds $80~$MeV in the CMS frame must be fewer than five.}
\item{For the four-vector of missing momentum defined by $p_{\mathrm{miss}}=p_{\mathrm{beam}}-p_{\mathrm{obs}}$,
 the missing mass $M_{\rm miss}$ defined as 
 $M^2_{\mathrm{miss}} = p_{\mathrm{miss}}^2$ must lie in the range $1$ GeV$/c^2$ $\leq M_{\mathrm{miss}}\leq 7$ GeV$/c^2$, where $p_{\rm beam}$ and $p_{\rm obs}$
 are the four-momentum of the beam and all detected particles, respectively.}
\item{The missing-momentum's polar angle must satisfy $30^{\circ}\leq \theta_{\mathrm{miss}} \leq 150^\circ$.}
\end{itemize}

\subsection{Final selection}
The candidates of the daughter particles of $\tau^+\tau^- \rightarrow (\pi^+ \pi^0 \bar{\nu})(l^-\nu\bar{\nu}\gamma )$,
 \textit{i.e.,} the lepton, photon, and charged and neutral pions, are assigned in each of the preselected events. 
\begin{itemize}
\item{The lepton candidate is selected using likelihood-ratio values.
 The electron selection uses $P_e = {L_e}/(L_e + L_x) > 0.9$, where $L_e$ and $L_x$ are the track's likelihood values for the electron and non-electron hypotheses, respectively.
 These values are determined using specific ionization ($\mydif E/\mydif x$) in the CDC, the ratio of ECL energy and CDC momentum $E/P$,
 the transverse shape of the cluster in the ECL, the matching of the track with the ECL cluster and the light yield in the ACC~\cite{citeEID}.
 The muon selection uses the likelihood ratio $P_\mu = {L_\mu}/(L_\mu+L_\pi+L_K)> 0.9$, where the likelihood values are
 determined by the measured vs expected range for the $\mu$ hypothesis
 and transverse scattering of the track in the KLM~\cite{citeMID}.
 The reductions of the signal efficiencies with lepton selections are approximately 2\% and 3\% for the electron and muon, respectively.
 The pion candidates are distinguished from kaons using $P_\pi = {L_\pi}/(L_\pi + L_K) > 0.4$, where the likelihood values are determined by the ACC response,
 the timing information from the TOF and $\mydif E/\mydif x$ in the CDC. The reduction of the efficiency with pion selection is approximately 5\%.
}
 
\item{The $\pi^0$ candidate is formed from two photon candidates, where each photon satisfies $E_\gamma > 80$~MeV, with an invariant mass of $115$~MeV$/c^2$ $<M_{\gamma\gamma}<150$~MeV$/c^2$.
 Figure~\ref{pi0dist} shows the distribution of the invariant mass of the $\pi^0$ candidates.
 The reduction of the signal efficiency is approximately 24\%.}
\item{The $\rho$ candidate is formed from a $\pi$ and a $\pi^0$ candidate, with an invariant mass of $0.5~\mathrm{GeV}/{c^2}<M_{\pi\pi^0}<1.5$~GeV$/{c^2}$.
 Figure~\ref{rhodist} shows the distribution of the invariant mass of the $\rho$ candidates.
 The reduction of the signal efficiency is approximately 3\%.}
\item{The signal photon candidate's energy must exceed $80$~MeV if within the ECL barrel ($31.4^\circ<\theta_\gamma < 131.5^\circ$) or
 $100$~MeV if within the ECL endcaps ($12.0^\circ<\theta_\gamma < 31.4^\circ$ or $131.5^\circ<\theta_\gamma < 157.1^\circ$).
 As shown in Fig.~\ref{selcoslg}, this photon must lie in a cone determined by the lepton-candidate direction that is defined by cos$\theta_{e\gamma}>0.9848$ and
 cos$\theta_{\mu\gamma}>0.9700$ for the electron and muon mode, respectively.
 The reductions of the signal efficiencies for the requirement
 on this photon direction are approximately 11\% and 27\% for electron and muon mode, respectively.
 Furthermore, if the photon candidate and either of the photons from the $\pi^0$, which is a daughter of the $\rho$ candidate, form an invariant mass of the
 $\pi^0$ ($115$~MeV$/c^2$ $<M_{\gamma\gamma}<150$~MeV$/c^2$), the event is rejected. The additional selection reduces the signal efficiencies by $1\%$.}
\item{The direction of the combined momentum of the lepton and photon in the CMS frame must not enter the hemisphere determined by the $\rho$ candidate:
 event should satisfy $\theta_{(l\gamma)\rho}>90^\circ$. This selection reduces the signal efficiency by $0.4\%$.
}
\item{There must be no additional photons in the aforementioned cone around the lepton candidate; the sum of the energy in the laboratory frame of
 all additional photons that are not associated
 with the $\pi^0$ or the signal photon (denoted as $E^\mathrm{LAB}_{\mathrm{extra}\gamma}$) should not exceed 0.2~GeV and 0.3~GeV for
 the electron and muon mode, respectively. The reductions of the signal efficiencies for the requirement
 on the $E^\mathrm{LAB}_{\mathrm{extra}\gamma}$ are approximately 14\% and 6\% for electron and muon mode, respectively.}
\end{itemize}

\begin{figure}[]
{\centering \subfloat[]{\includegraphics[width=8.0cm]{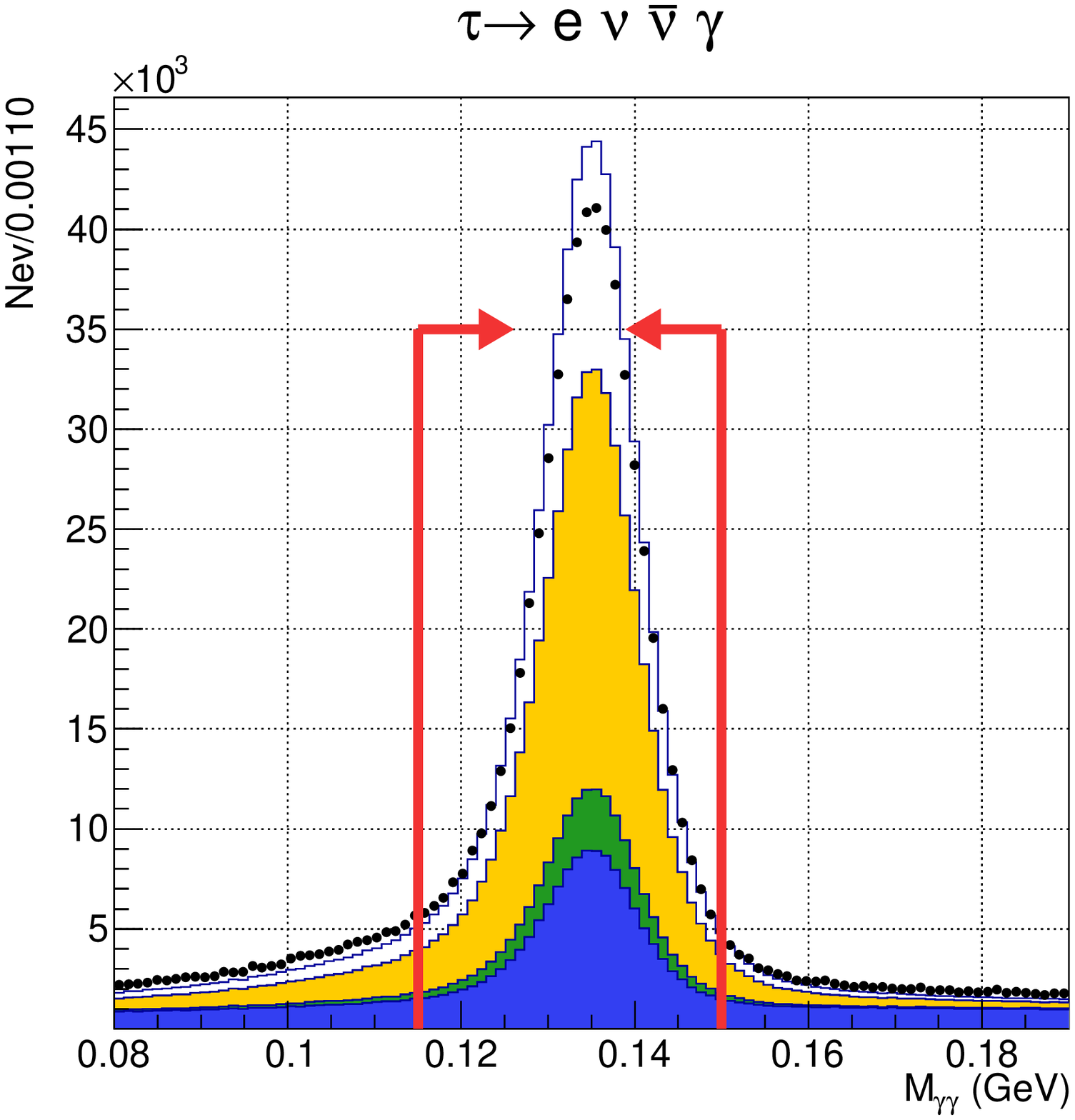}\label{pi0dist:pi0e}} }
{\centering \subfloat[]{\includegraphics[width=8.0cm]{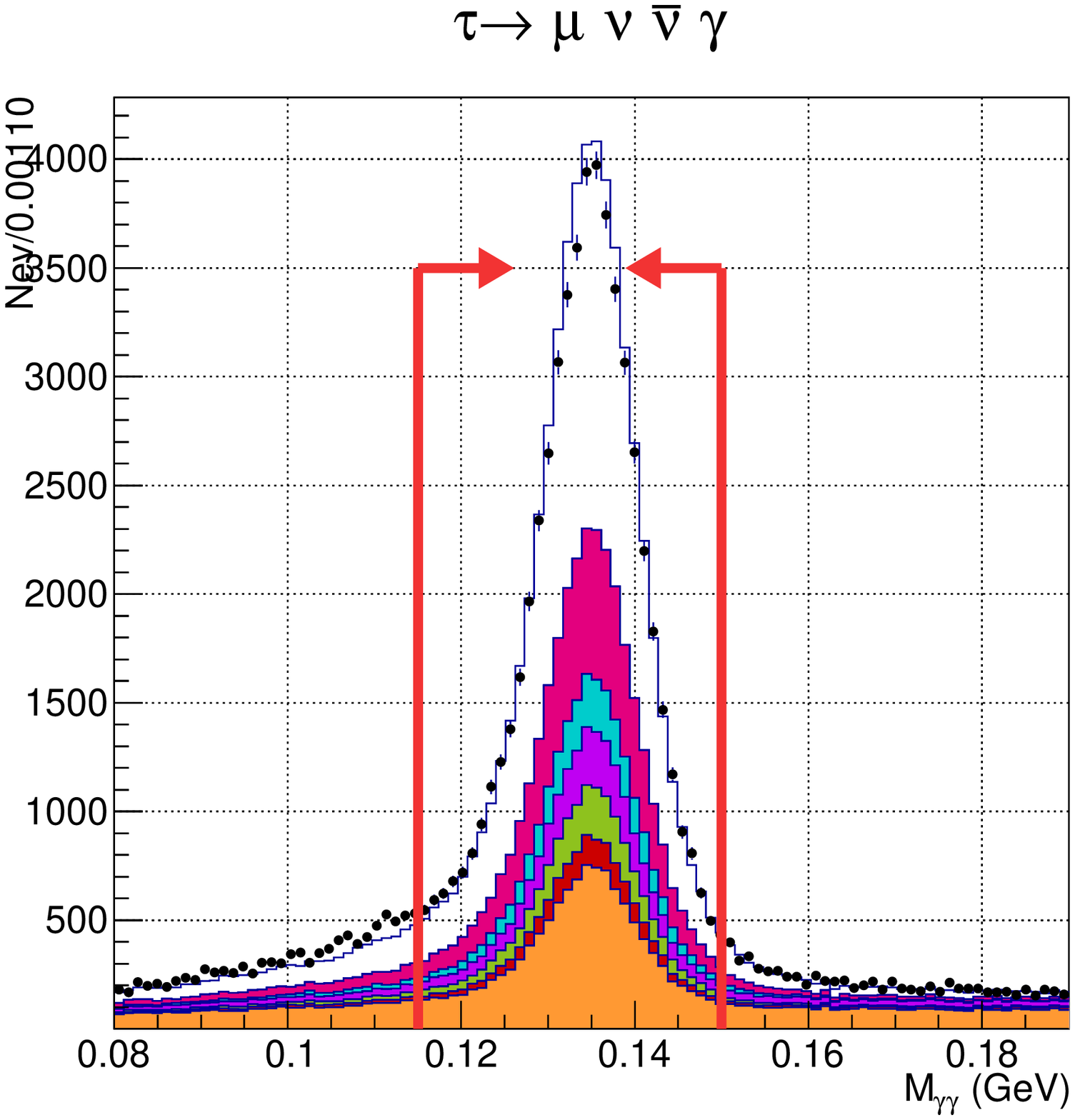}\label{pi0dist:pi0mu}}} 
\caption[]{\raggedright  Distribution of $M_{\gamma \gamma}$. Dots with error bars are experimental data and histograms are MC distributions.
 The MC histograms are scaled to that of experimental one based on the numbers just after the preselection.
 The red arrows indicate the selection window $115$~MeV$/c^2$ $<M_{\gamma\gamma}<150$~MeV$/c^2$.
 The reduction of the efficiency is approximately 24\% for both electron and muon modes.
 
 (a) $\tau \rightarrow e \nu\bar{\nu}\gamma$ candidates:
 the open histogram corresponds to the signal, the yellow and green histograms represent ordinary leptonic decay plus bremsstrahlung and radiative leptonic decay plus bremsstrahlung,
 respectively, and the blue histogram represents other processes.
 
 (b) $\tau \rightarrow \mu \nu\bar{\nu}\gamma$ candidates: the open histogram corresponds to signal, the magenta histogram represents ordinary leptonic decay plus beam background,
 the water-blue histogram represents ordinary leptonic decay plus ISR/FSR processes, the purple histogram represents three-pion events where
 $\tau^+ \rightarrow\pi^+\pi^0\pi^0 \bar{\nu}$ is misreconstructed as a tagging $\tau^+\rightarrow \pi^+\pi^0 \bar{\nu}$ candidate, the green histogram represents $\rho$-$\rho$ background
 where $\tau^- \rightarrow \pi^-\pi^0\nu$ is selected due to misidentification of pion as muon, the red histogram represents 3$\pi$-$\rho$ events
 where $\tau^- \rightarrow \pi^-\pi^0\pi^0\nu$ is selected by misidentification similarly to the $\rho$-$\rho$ case and the orange histogram represents other processes.
 }\label{pi0dist}
\end{figure}

\begin{figure}[]
{\centering \subfloat[]{\includegraphics[width=8.0cm]{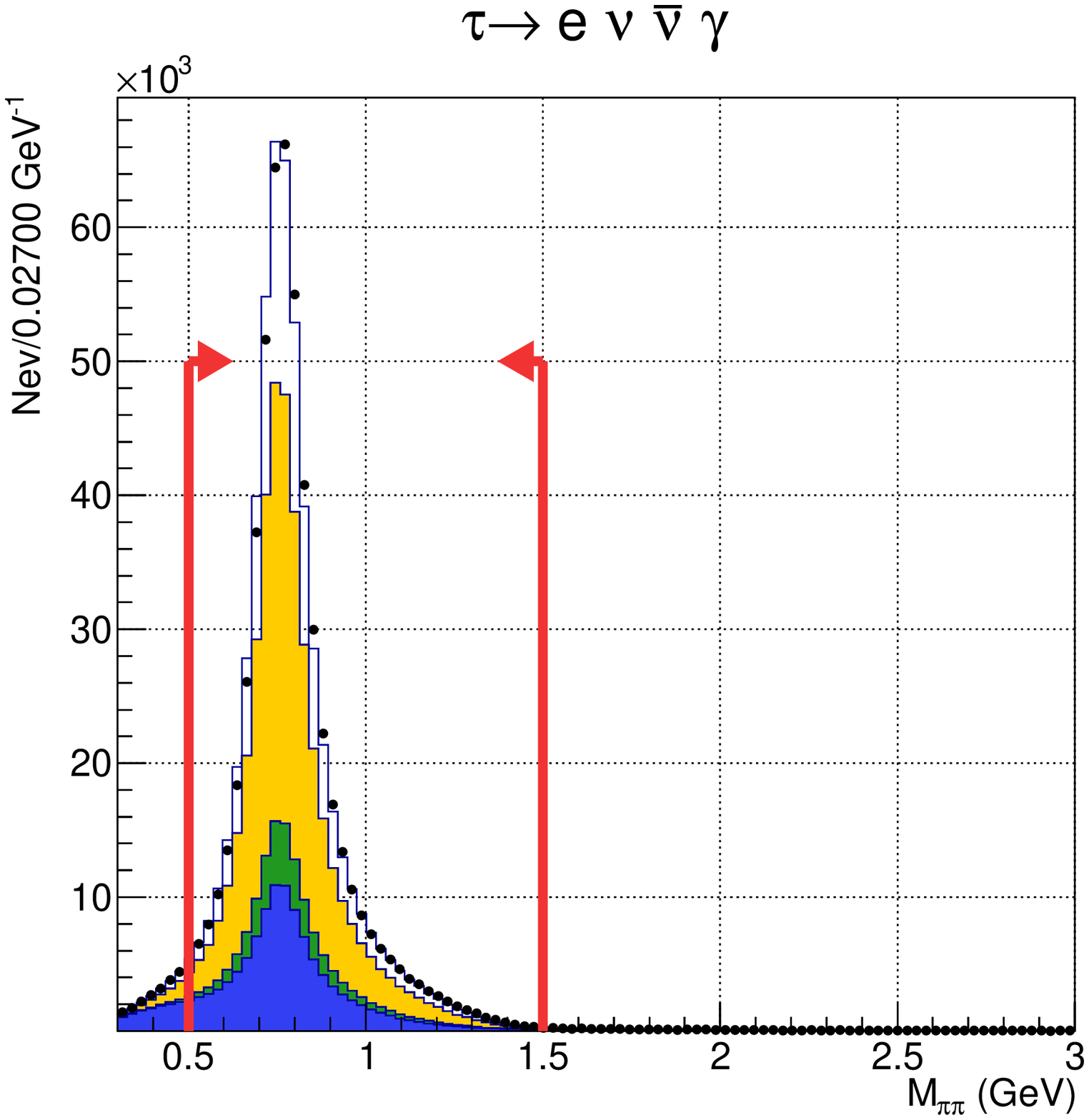}\label{pi0dist:rhoe}} }
{\centering \subfloat[]{\includegraphics[width=8.0cm]{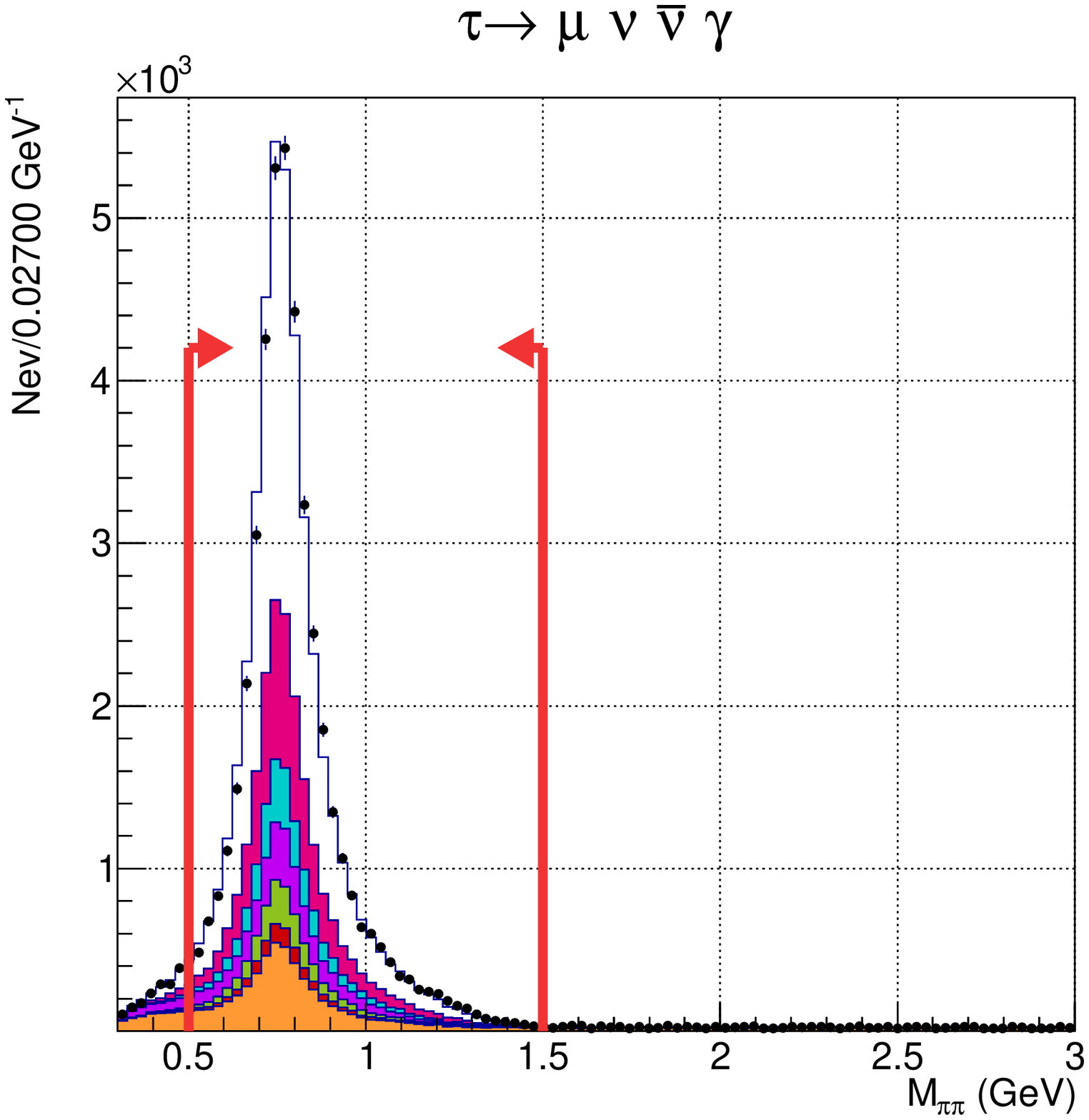}\label{pi0dist:rhomu}}}
\caption[]{\raggedright  Distribution of $M_{\pi \pi^0}$:
 (a) $\tau \rightarrow e \nu\bar{\nu}\gamma$ candidates and (b) $\tau \rightarrow \mu \nu\bar{\nu}\gamma$ candidates.
 Dots with error bars are experimental data and histograms are MC distributions.
 The color of each histogram is explained in Fig.~\ref{pi0dist}.
 The red arrows indicate the selection window $0.5$~GeV$/c^2$ $<M_{\pi\pi^0}<1.5$~MeV$/c^2$.
 The reduction of the efficiency is approximately 3\% both for electron and muon modes.}\label{rhodist}
\end{figure}

\begin{figure}[]
{\centering \subfloat[]{\includegraphics[width=8.0cm]{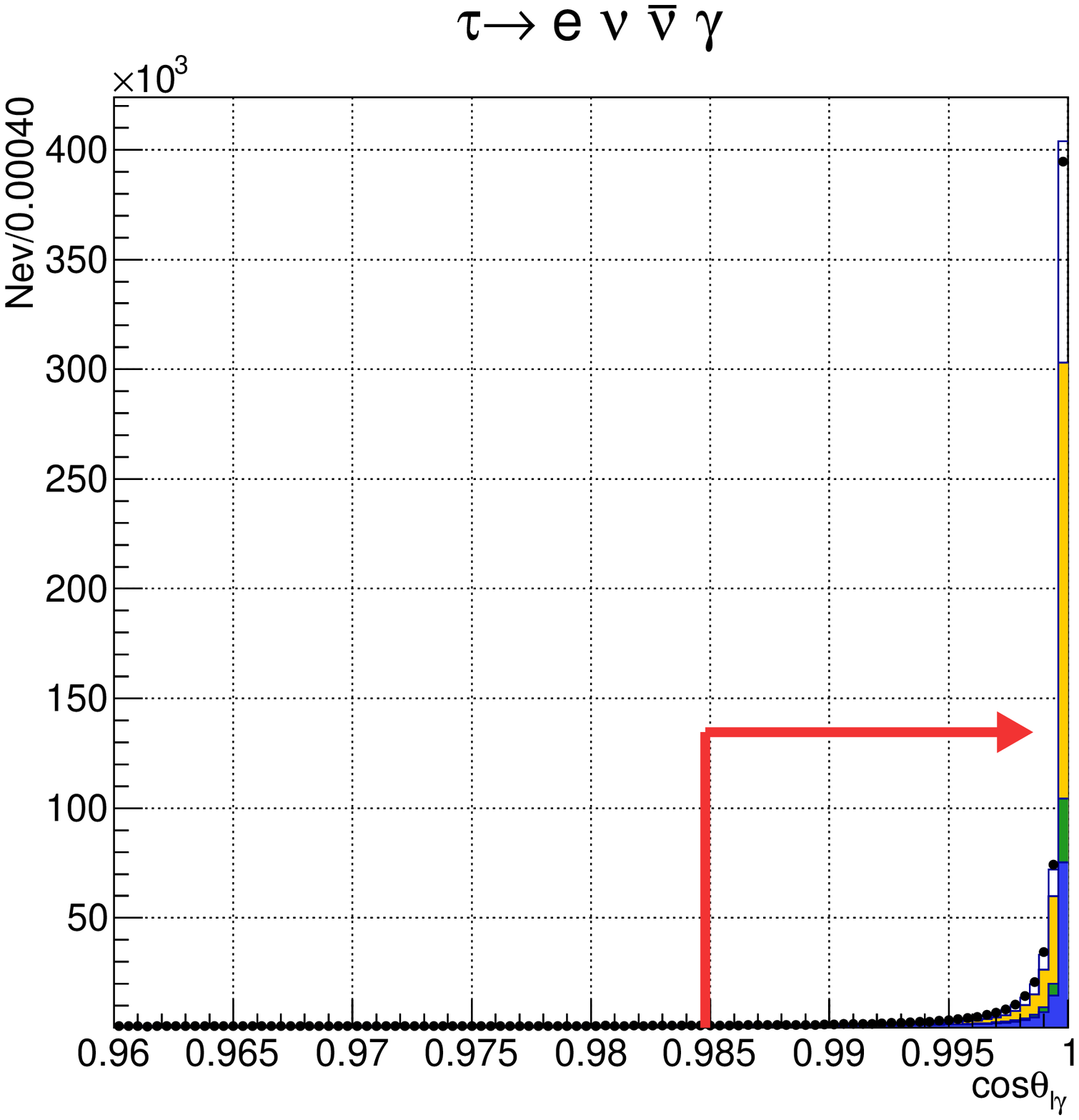}\label{selcoslg:e}} }
{\centering \subfloat[]{\includegraphics[width=8.0cm]{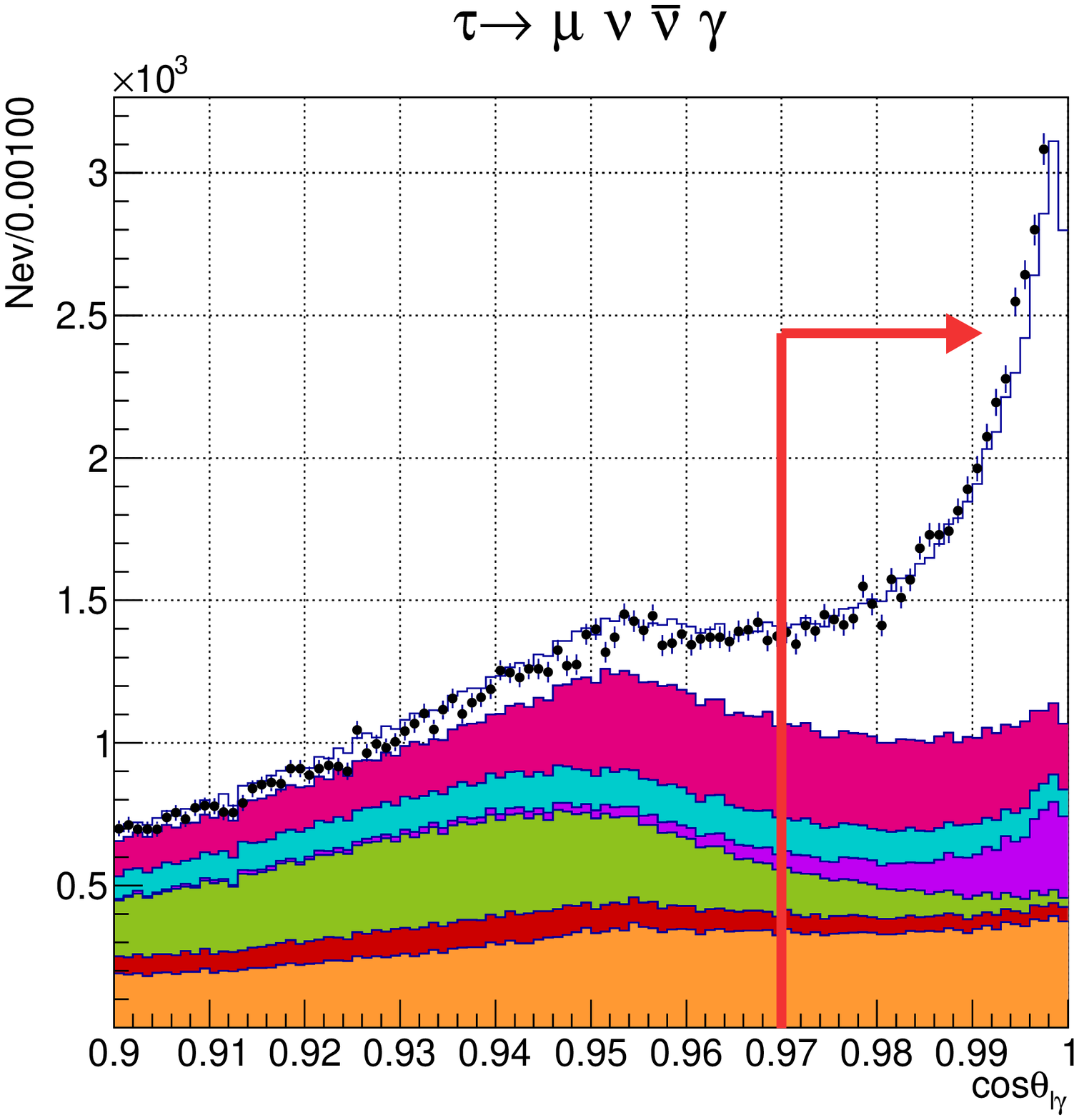}\label{selcoslg:mu}}} 
\caption[]{\raggedright  Distribution of $\cos\theta_{l\gamma}$:
 (a) $\tau \rightarrow e \nu\bar{\nu}\gamma$ candidates and (b) $\tau \rightarrow \mu \nu\bar{\nu}\gamma$ candidates.
 Dots with error bars are experimental data and histograms are MC distributions.
 The color of each histogram is explained in Fig.~\ref{pi0dist}.
 The red arrows indicate the selection condition cos$\theta_{e\gamma}>0.9848$ and
 cos$\theta_{\mu\gamma}>0.9700$ for the electron and muon mode, respectively.
 The reduction of the efficiency is approximately 11\% and 27\% for electron and muon modes.}\label{selcoslg}
\end{figure}

These selection criteria are optimized using a MC method
 where $e^+e^-\rightarrow \tau^+ \tau^-$ pair production and the successive decay of the $\tau$
 are simulated by KKMC~\cite{citeKKMC} and TAUOLA~\cite{citeTAUOLA1,citeTAUOLA2}
 generators, respectively. The detector effects are simulated based on the GEANT3 package~\cite{citegeant3}.

\begin{figure}[]
{\centering \subfloat[$E_\gamma$]{\includegraphics[width=8.0cm]{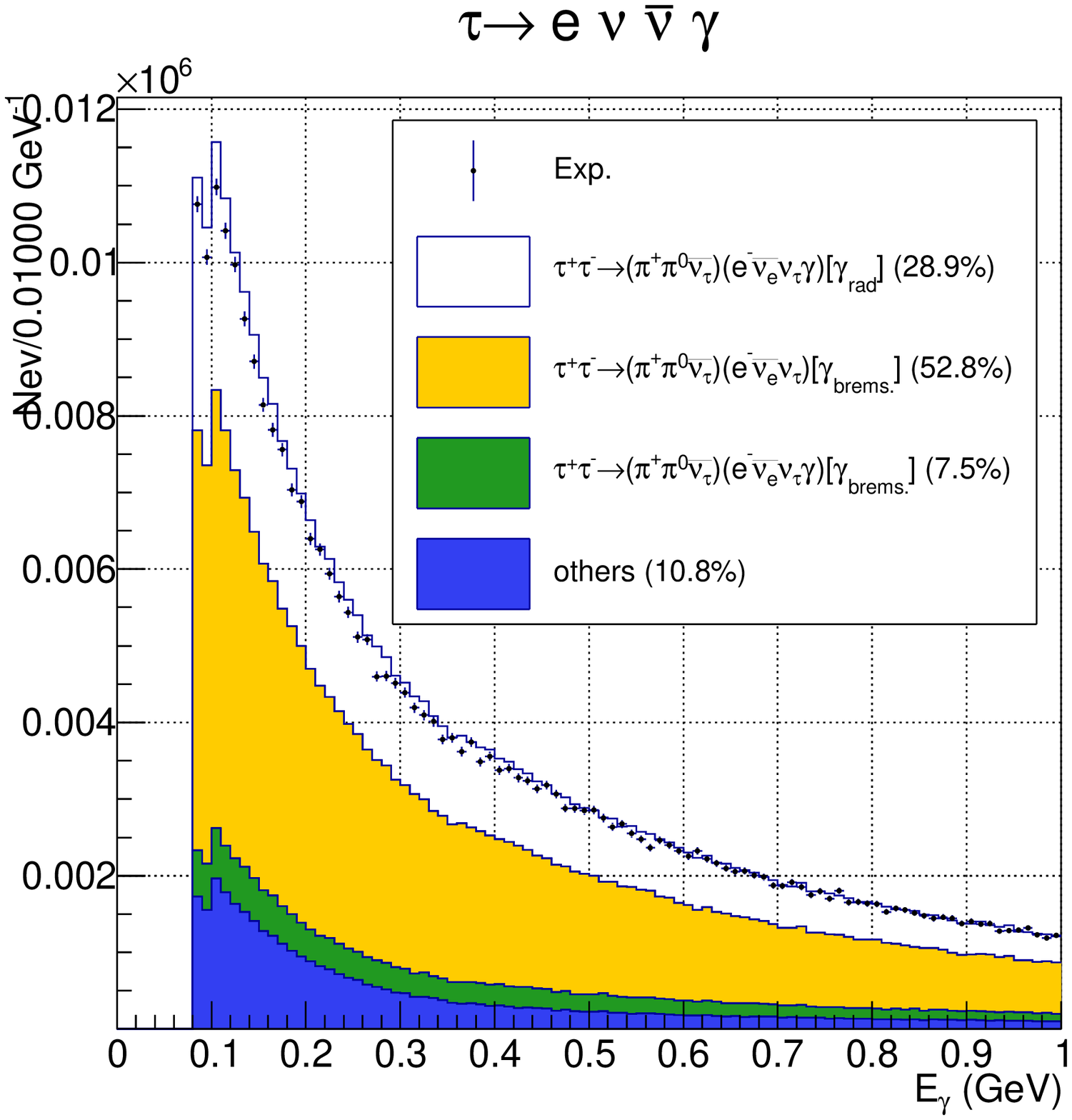}\label{selection:E_e}} }
{\centering \subfloat[$\theta_{e\gamma}$]{\includegraphics[width=8.0cm]{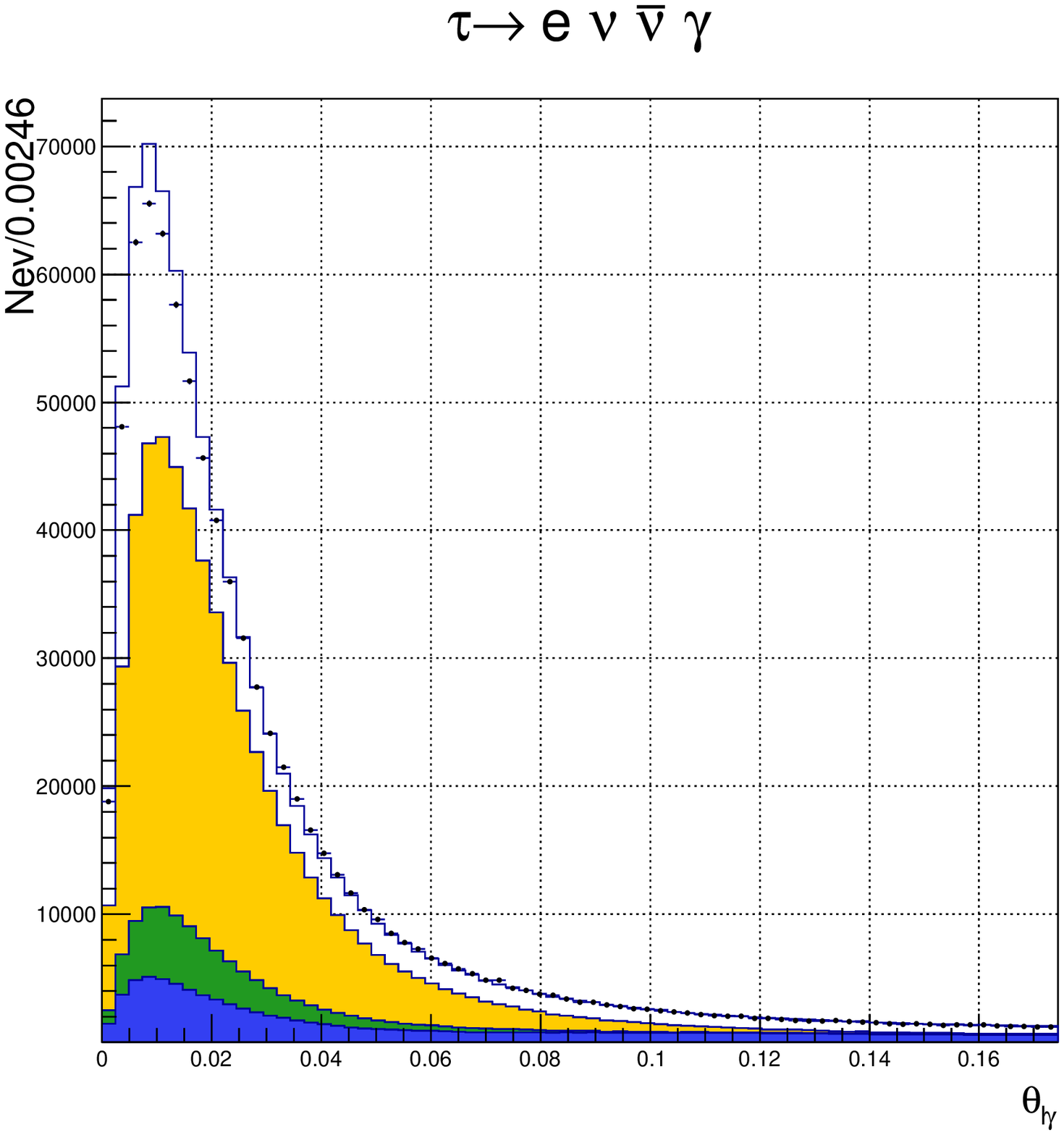}\label{selection:psi_e}}} \\
\caption[]{\raggedright  Final distribution of (a) photon energy $E_\gamma$ and (b) $\theta_{e\gamma}$
 for the $\tau^+\tau^- \rightarrow (\pi^+ \pi^0 \bar{\nu})(e^- \nu \bar{\nu} \gamma)$ decay candidates.
 Dots with error bars are experimental data and histograms are MC distributions.
 The color of each histogram is explained in Fig.~\ref{pi0dist}.
  }\label{comp_rem_e}
\end{figure}
\begin{figure}[]
{\centering \subfloat[$E_\gamma$]{\includegraphics[width=8.0cm]{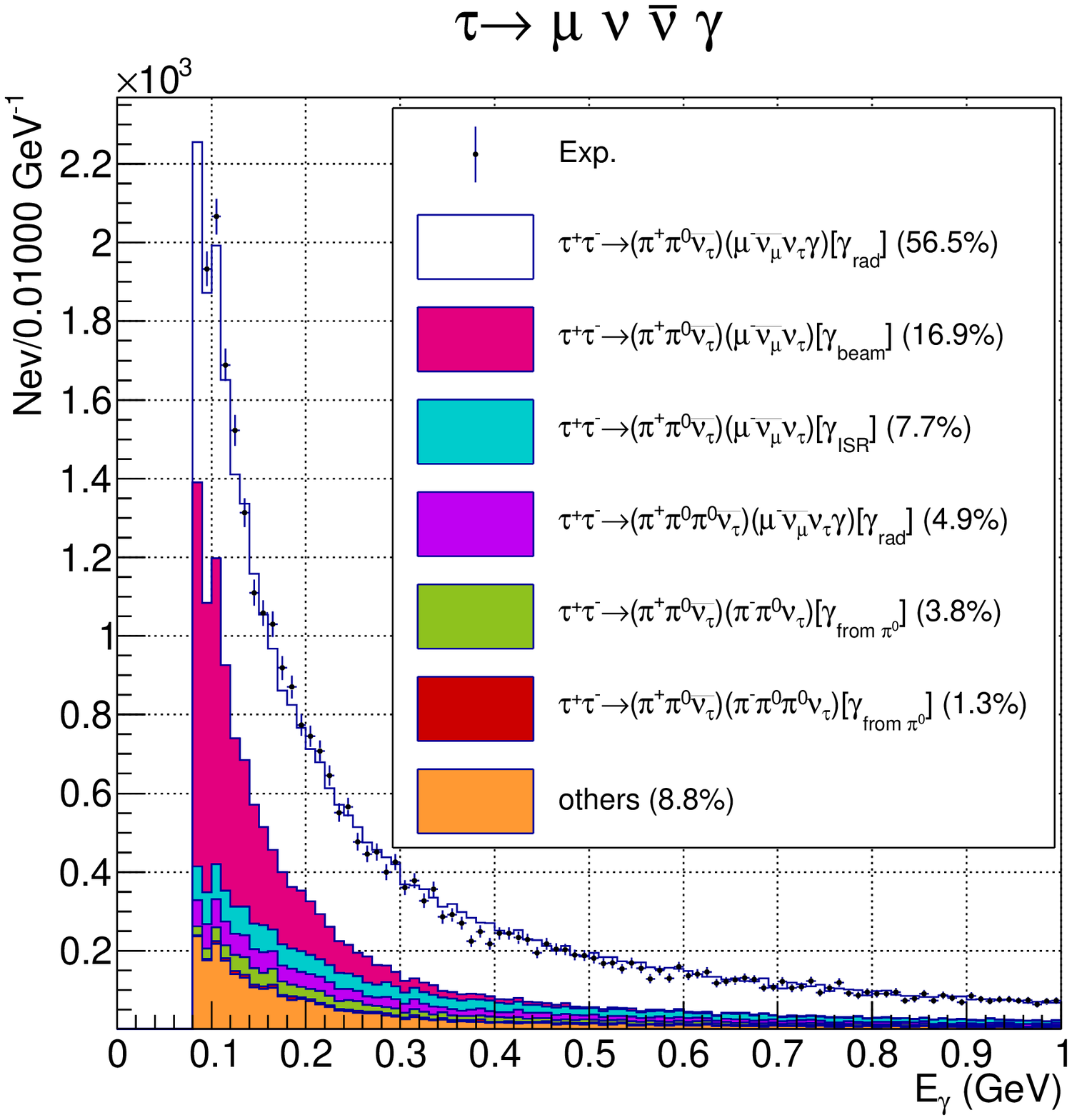}\label{selection:E_mu}} }
{\centering \subfloat[cos$\theta_{\mu\gamma}$]{\includegraphics[width=8.0cm]{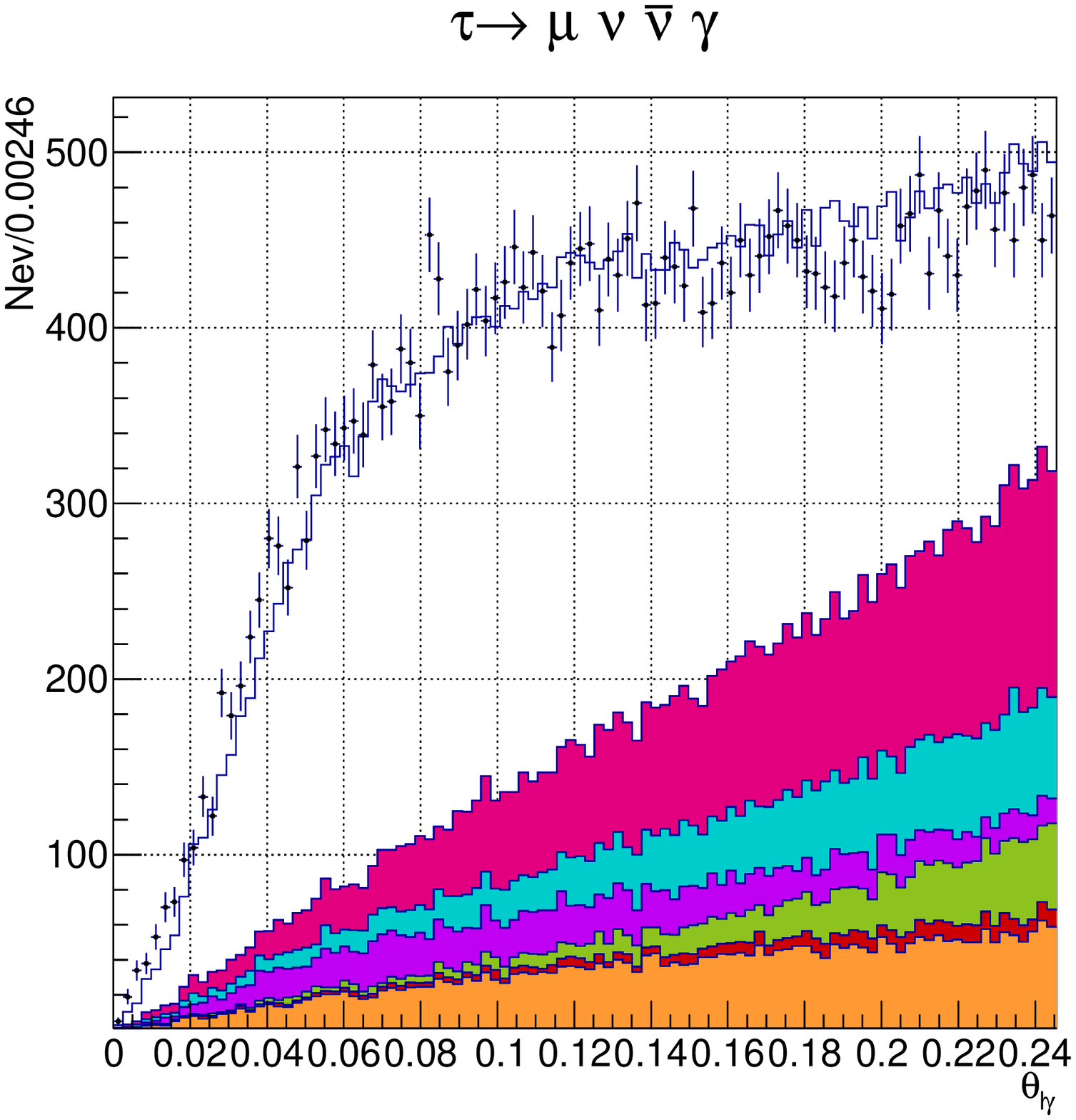}\label{selection:psi_mu}}}
\caption[]{\raggedright Final distribution of (a) photon energy $E_\gamma$ and (b) $\theta_{\mu\gamma}$~for~the~$\tau^+\tau^- \rightarrow (\pi^+\pi^0 \bar{\nu})(\mu^- \nu \bar{\nu} \gamma)$ decay candidates.
 Dots with error bars are experimental data and histograms are MC distributions.
 The color of each histogram is explained in Fig.~\ref{pi0dist}.
 }\label{remgraphs_mu}
\label{comp_rem_mu}
\end{figure}

The selection criteria suppress
background while retaining efficiency for signal events.
 A characteristic feature of the radiative decay is that the photon tends
to be produced nearly collinear with the final-state lepton.
Distributions of the photon energy $E_\gamma$ and the angle between the lepton and photon $\theta_{l\gamma}$ for the selected events are shown in
Figs.~\ref{comp_rem_e} and \ref{comp_rem_mu} for $\tau^- \rightarrow e^- \nu \bar{\nu} \gamma$ and $\tau^- \rightarrow \mu^- \nu \bar{\nu} \gamma$ candidates, respectively.
 
In the electron mode, the fraction of the signal decay in the selected sample
 is about $30\%$ due to the large external bremsstrahlung rate in the non-radiative leptonic
 $\tau$ decay events. In the muon mode, the
fraction of the signal decay is about $60\%$; here, the main background arises from ordinary leptonic
 decay ($\tau^-\rightarrow l^- \nu \bar{\nu}$) events
where either an additional photon is reconstructed from a beam background
in the ECL or a photon is emitted by the
initial-state $e^+e^-$.

With the described selection criteria, the average efficiencies of signal events are evaluated by MC.
 The selection information is summarized in Table~\ref{evtsel_sum}.

\begin{table}[]
\caption{Summary of event selection \label{evtsel_sum}}
\begin{center}
\begin{tabular}{l|cccc} \hline \hline
Item & $(e^- \nu\bar{\nu}\gamma)(\pi^+\pi^0\bar{\nu})$ & $(e^+\nu\bar{\nu}\gamma )(\pi^-\pi^0{\nu})$ & $(\mu^-\nu\bar{\nu}\gamma )(\pi^+\pi^0\bar{\nu})$ & $(\mu^+\nu\bar{\nu}\gamma )(\pi^-\pi^0{\nu})$ \\ \hline
$N_{\mathrm{sel}}$ & 420005  & 412639 & 35984 & 36784 \\ 
$\varepsilon^{\dagger}$ (\%) & $4.45\pm0.19$ & $4.43\pm0.19$ & $3.42\pm0.15$  & $3.39\pm0.15$ \\
Purity (\%) & \multicolumn{2}{c}{$28.9$} & \multicolumn{2}{c}{$56.5$} \\ \hline \hline
\end{tabular}
\end{center}
\begin{flushleft}\vspace{-4mm}$~~~~~~~~^\dagger$~{\small The signal is defined by the photon energy threshold in the $\tau$-rest frame with $E_{\gamma}^{*}>10$~MeV.}\end{flushleft}
\end{table}

\section{Analysis}
 
Accounting for the event-selection criteria and the contamination from identified backgrounds, the total visible (properly normalized) PDF
 for the observable $\bvec{x}$ in each event is given by
\begin{equation}
P(\bvec{x}) = (1-\sum_{i}{\lambda_{i}}) \frac{S(\bvec{x})\varepsilon (\bvec{x})}{\int{\mathrm{d}\bvec{x} S(\bvec{x})\varepsilon (\bvec{x})}}
+\sum_{i}{\lambda_{i}\frac{B_{i}(\bvec{x})\varepsilon (\bvec{x})}{\int{\mathrm{d}\bvec{x} B_{i}(\bvec{x})\varepsilon (\bvec{x})}}},\label{generalpdf}
\end{equation}
where $S(\bvec{x})$ is the signal distribution given by Eq.~\ref{totform}, $B_{i}(\bvec{x})$
 is the distribution of the $i^{\rm th}$ category of background
 ($i$ runs $1,2,3$ and $1,2,\ldots,6$ for electron and muon modes, respectively and this index indicates each category
 filled with a color in Figs.~\ref{comp_rem_e} and \ref{comp_rem_mu}), $\lambda_{i}$ is the
fraction of this background and $\varepsilon (\bvec{x})$ is the
selection efficiency of signal distribution. In general, the $\varepsilon (\bvec{x})$ is not common between the signal and backgrounds,
 the difference, however, is included in the definition of $B_{i}(\bvec{x})$. The PDFs of the major background modes
 are basically described using their theoretical formulae while other minor contributions
 are treated as one category and described based on the MC simulation.

 In the integration of the differential cross section over the $\Phi$ in Eq.~\ref{totformwithJ},
 we randomly generate integration variables and calculate an average of the integrand.
 Moreover, an unfolding of the detector resolution is also performed in this integration,
 where the distortion of the momenta of the charged tracks and photon energies due to the finite accuracy of the detector
 is taken into account as a convolution with its resolution function determined from an MC simulation of the detector.
 When the generated kinematic variables are outside allowed phase space of the signal distribution, we assign zero to
 the integrand rather than using its unphysical (negative) value.
 This means that we discard trials which have negative-mass neutrinos.
 If such discarded trials in the integration exceed 20\% of the total number of iterations, we further reject the event.
 This happens for events which lie around the kinematical boundary of the signal phase space.
 The corresponding reduction of the efficiency is 2\% and 3\% for the electron and muon mode, respectively.
 This additional decrease of the efficiency is not reflected on the values of Table~\ref{evtsel_sum}.

From $P(\bvec{x})$, the negative logarithmic
likelihood function (NLL) is constructed and the best estimators of the
Michel parameters, $\bar{\eta}$ and $\xi\kappa$, are obtained by
minimizing the NLL. The efficiency $\varepsilon(\bvec{x})$ is a common
multiplier in Eq.~\ref{generalpdf} and does not depend on the Michel
parameters. This is one
of the essential features of the unbinned maximum likelihood method that we
use. We validated our fitter and procedures using a MC sample generated according to the SM distribution.
 The fitted results are consistent with the SM predictions within 1$\sigma$ statistical deviation of the experimental result. 

\subsection{Analysis of experimental data }

The analysis of experimental data is performed in the same way as MC simulation.
 The difference between real data and MC simulation is represented by the measured correction factor
 $R(\bvec{x})=\varepsilon_{\mathrm{ex}}(\bvec{x})/\varepsilon_{\mathrm{MC}}(\bvec{x})$ that is close to unity; its extraction is described below.
 With this correction, Eq.~\ref{generalpdf} is modified to
\begin{equation}
P^{\mathrm{ex}}(\bvec{x}) = (1-\sum_{i}{\lambda_{i}})\cdot \frac{S(\bvec{x})\varepsilon_{\mathrm{MC}} (\bvec{x})R(\bvec{x})}{\int{\mathrm{d}\bvec{x} S(\bvec{x})\varepsilon_{\mathrm{MC}} (\bvec{x}) R(\bvec{x}) }}
+\sum_{i}{\lambda_{i}\frac{B_{i}(\bvec{x})\varepsilon_{\mathrm{MC}} (\bvec{x}) R(\bvec{x}) }{\int{\mathrm{d}\bvec{x} B_{i}(\bvec{x})\varepsilon_{\mathrm{MC}} (\bvec{x})} R(\bvec{x}) }}\label{generalpdfR}.
\end{equation}
The presence of $R(\bvec{x})$ in the numerator does not affect the NLL minimization, but its presence in the denominator does.

We evaluate $R(\bvec{x})$ as the product of the measured corrections for the trigger,
 particle identifications and track reconstruction efficiencies:
\begin{align}
&R(\bvec{x})=R_\mathrm{trg}R_l(P_l, \cos\theta_l)R_{\gamma }(P_\gamma, \cos\theta_\gamma)R_\pi(P_\pi, \cos\theta_\pi)R_{\pi^0}(P_{\pi^0}, \cos\theta_{\pi^0}), \\
&R_l(P_l, \cos\theta_l) = R_\mathrm{rec.}(P_l, \cos\theta_l) R_\mathrm{LID}(P_l, \cos\theta_l), \\
&R_\pi (P_\pi, \cos\theta_\pi) = R_\mathrm{rec.}(P_\pi, \cos\theta_\pi) R_\mathrm{\pi ID}(P_\pi, \cos\theta_\pi ).
\end{align}

The lepton identification efficiency correction is estimated using the two-photon processes $e^+e^-\rightarrow e^+e^-l^+l^-$ ($l=e$ or $\mu$).
 Since the momentum of the lepton from the two-photon process ranges from $0$ to approximately $4$~GeV$/c$ in the laboratory frame,
 the efficiency correction factor can be evaluated for our signal process as a function of $P_l$ and $\cos\theta_l$.

The track reconstruction efficiency correction is extracted from $\tau^+\tau^- \rightarrow (l^+\nu\bar{\nu})(\pi^-\pi^+\pi^- {\nu})$ events.
 Here, we count the number of events $N_4$ ($N_3$) in which four (three) charged tracks are reconstructed.
 The three-charged-track event is required to have a negative net charge ($\pi^+$ is missing).
 Since the charged track reconstruction efficiency $\varepsilon$ is included as, respectively,
 $\varepsilon^4$ and $\varepsilon^3(1-\varepsilon)$ in $N_4$ and $N_3$,
 the value of $\varepsilon$ can be obtained by $\varepsilon = N_4/(N_4+N_3)$.

 The $\pi^0$ reconstruction efficiency correction is obtained by comparing the ratio of the number
 of selected events of $\tau^+\tau^- \rightarrow (\pi^+ \pi^0 \bar{\nu})(\pi^- \pi^0 \nu)$ and $\tau^+\tau^- \rightarrow (\pi^+ \pi^0 \bar{\nu})(\pi^- \nu)$ between
 experiment and MC simulation.
 If we ignore one of the photon daughters from the $\pi^0$, the $\gamma$ reconstruction efficiency correction can be also extracted in the same manner.
 
 The trigger efficiency correction has the largest impact among these factors.
 In particular, for the electron mode, because of the similar structure of our signal events
 and Bhabha events (back-to-back topology of two-track events),
 many signal events are rejected by the Bhabha veto in the trigger.
 The veto of the trigger results in a spectral distortion and a large systematic uncertainty.
 The correction factor from trigger is tabulated using
 the charged and neutral triggers (denoted as $Z$ and $N$), which provide completely independent signals.
 Since the trigger is fired when both signals are not inactive,
 its efficiency is given by $\varepsilon_{\rm trg}= 1-(1-\varepsilon_Z)(1-\varepsilon_N)$,
 where $\varepsilon_Z$ and $\varepsilon_N$ are,
 respectively, the efficiencies of the charged and
 neutral triggers. Here, each efficiency is extracted by a comparison of the numbers $N_{Z\&N}/N_{N}$ or $N_{Z\&N}/N_{Z}$.
 Thus $R_{\rm trg}$ is obtained as a ratio of $\varepsilon_{\rm trg}$ between the experiment and MC simulation. Through the $R_{\rm trg}$, we incorporate this systematic bias and its uncertainty into our results. 
 
 Figure~\ref{corr_figs} shows the distribution of the momentum and the cosine of the polar angle of electron and muon events.
 In the figure, the effects of all corrections are seen mainly at cos$\theta_e<-0.6$ and cos$\theta_\mu<-0.6$.
\begin{figure}[]
{\centering \subfloat[$P_e$]{\includegraphics[width=7.2cm]{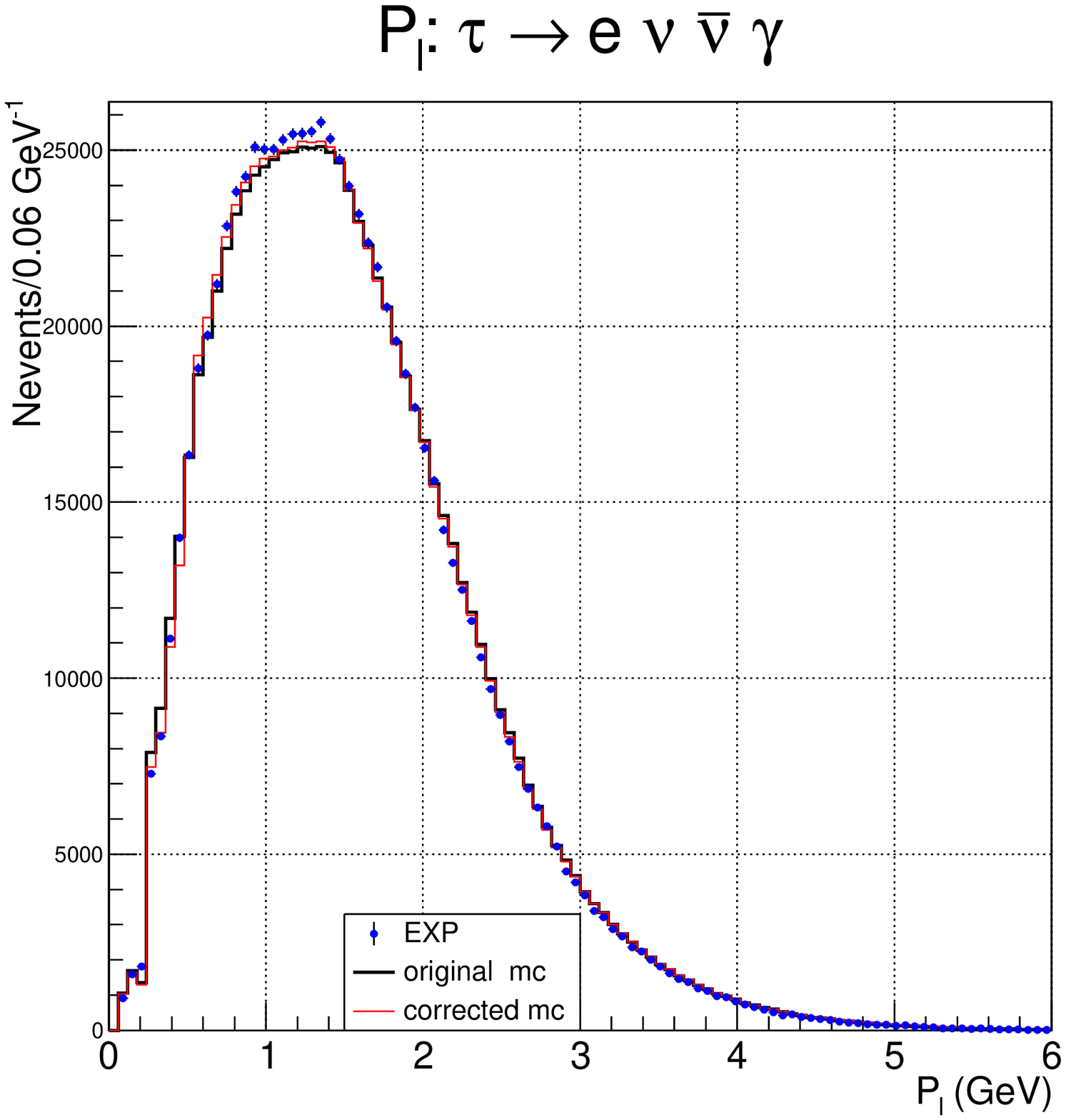}\label{selection:p_e}} }
{\centering \subfloat[cos$\theta_{e}$]{\includegraphics[width=7.2cm]{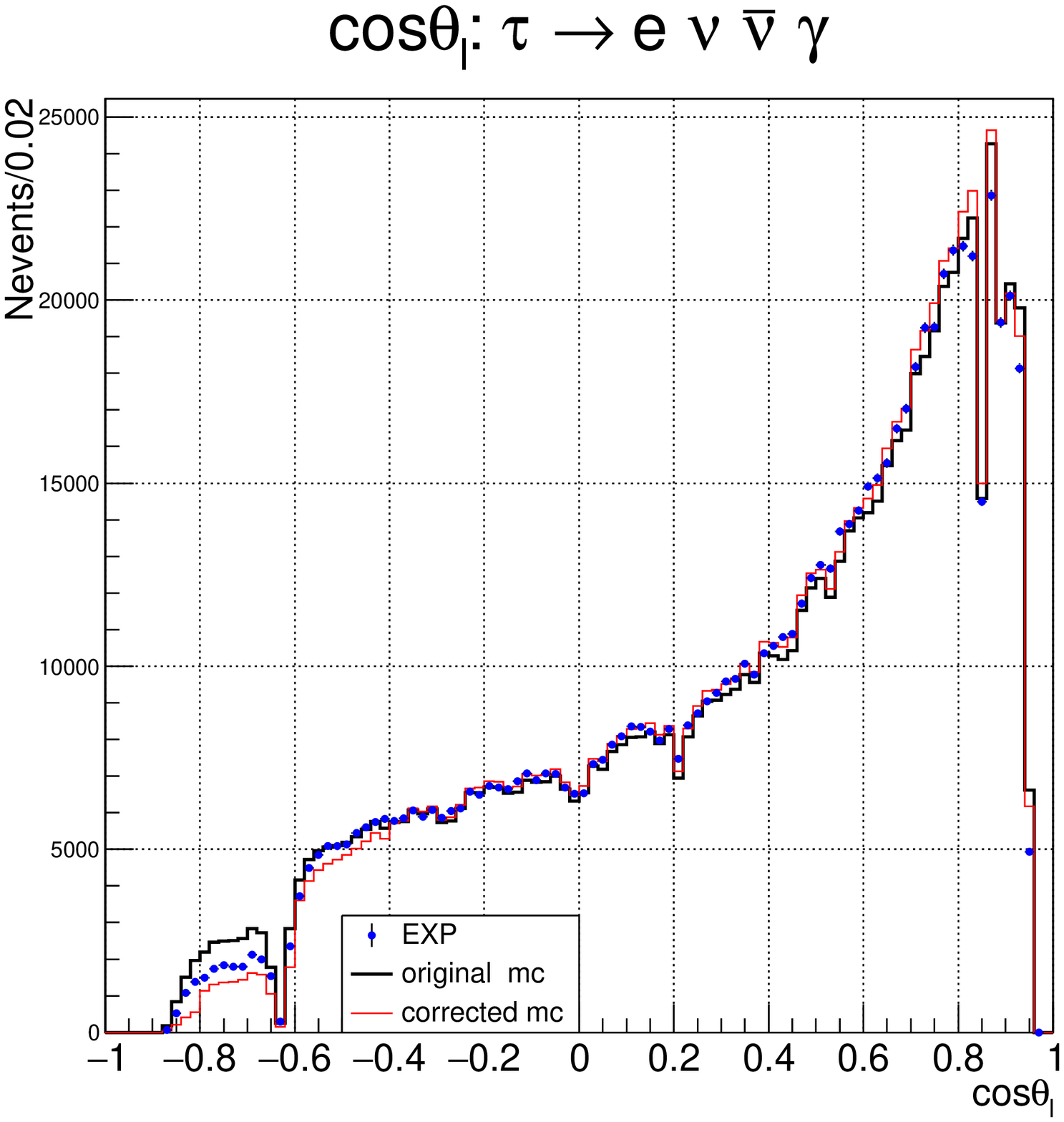}\label{selection:cth_e}} }\\
{\centering \subfloat[$P_\mu$]{\includegraphics[width=7.2cm]{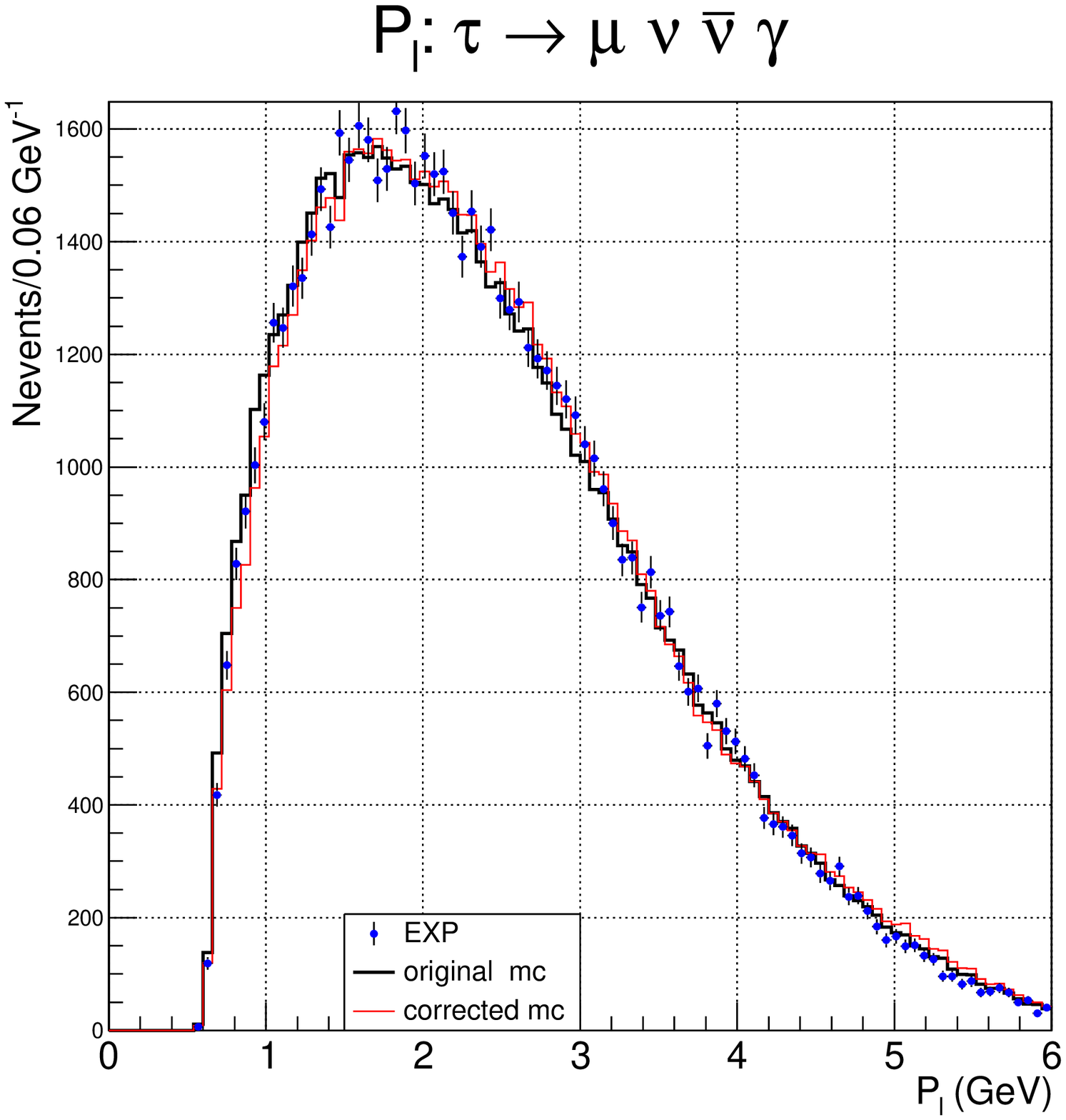}\label{selection:p_mu}} }
{\centering \subfloat[cos$\theta_{\mu}$]{\includegraphics[width=7.2cm]{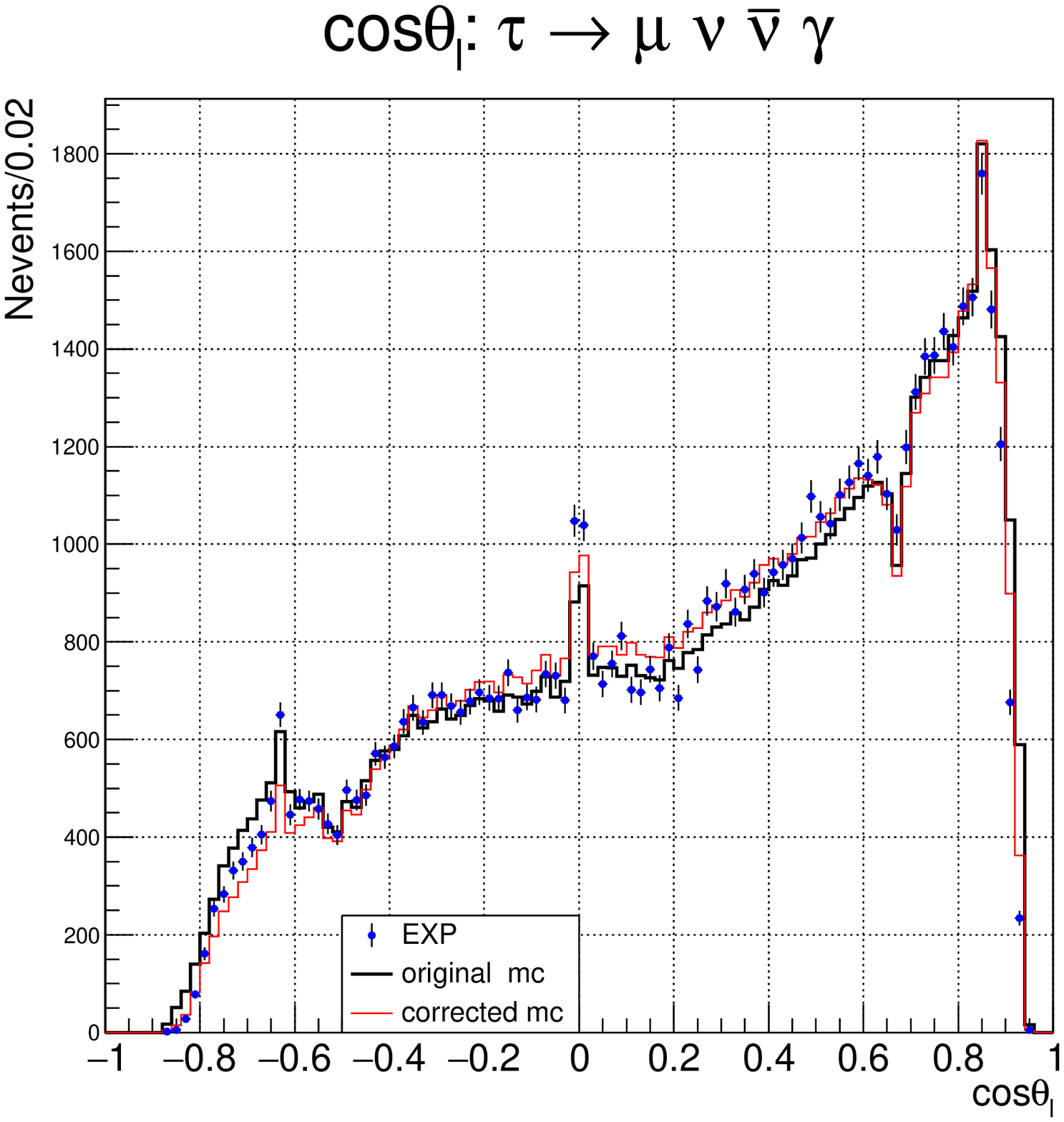}\label{selection:cth_mu}}}
\caption[]{\raggedright Distribution of (a)(c) momentum and (b)(d) cosine of the lepton direction for (a)(b) $\tau \rightarrow e \nu \bar{\nu} \gamma$
 and (c)(d) $\tau \rightarrow \mu \nu \bar{\nu} \gamma$ candidates.
 Blue points with error bars represent the experimental data. The black and red lines represent the distributions of the original and corrected MC, respectively. }\label{corr_figs}
\end{figure}

\subsection{Evaluation of systematic uncertainties}

In Table~\ref{sys_sum}, we summarize the contributions of various sources of systematic uncertainties.
 The dominant systematic source for the electron mode is the calculation of the relative normalizations.
 Due to the peculiarity of the signal PDF when $m_l\rightarrow 0$, the convergence of the factor is quite
 slow and results in a notable effect.
 The uncertainty of the relative normalization is evaluated using the central limit theorem.
 For a given number of MC events $N$,
 the errors of the relative normalizations $\left< A_i/A_0 \right>$ ($i=1,2$) are
 evaluated by $\sigma^2 = {\rm Var}(A_i/A_0)/N$, where ${\rm Var(X)}$ represents the variance of a random variable $X$. 
 The resulting systematic effect on the Michel parameter is estimated by varying the normalizations.
 The effect of the absolute normalization is estimated in the same way.
 
 The largest systematic uncertainty for the muon mode is due to the limited precision of the description
 of the background PDF that appears in Eq.~\ref{generalpdfR}.
 As mentioned before, the set of minor sources is treated as one additional category that is based on MC distributions.
 This effective description can generally discard information about correlations
 in the phase space and thereby give significant bias.
 The residuals of the fitted Michel parameters from the SM prediction obtained by the fit to the MC distribution
 are taken as the corresponding systematic uncertainties. 

Other notable uncertainties come from the limited knowledge of the measured branching ratios.
In particular, the uncertainties of the branching ratio of the radiative decay $\tau^-\rightarrow l^- \nu \bar{\nu} \gamma$ dominate the contribution.
The systematic effects of the cluster-merge algorithm in the ECL are evaluated as a function of the
 angle between the photon and lepton clusters at the ECL's front face ($\theta_{l\gamma}^{\mathrm{ECL}}$).
 The limit $\theta_{l\gamma}^{\mathrm{ECL}} \rightarrow 0$ represents the merger of the two clusters
 and the comparison of the distribution between experiment and MC gives us the corresponding bias.
 Detector resolutions of the photon energies and track momenta are evaluated by
 comparing the results obtained with and without the unfolding of the measured values.
 The error of the tabulated correction factor $R$ is estimated by varying the central values based on the uncertainty in each bin.
 The effect of the beam-energy spread is estimated by varying the input of this
 value for the calculation of PDF with respect to run-dependent uncertainties.
\begin{table}[]
\caption{List of systematic uncertainty contributions \label{sys_sum}}
\begin{center}
\begin{tabular}{lcccc} \hline \hline
Item & $\sigma_{\bar{\eta }}^e$ & $\sigma_{\xi \kappa }^e$ & $\sigma_{\bar{\eta }}^\mu$ & $\sigma_{\xi \kappa }^\mu$ \\ \hline
Relative normalizations & $4.2$ & $0.94$  & 0.15    & 0.04  \\
Absolute normalizations & $1.0$ & $0.01$ &  $0.03$ & $0.001$ \\
Description of the background PDF & $2.5$ & $0.24$ & 0.67 & 0.22  \\
Input of branching ratio & $3.8$ & $0.05$ & $0.25$ & $0.01$ \\
Effect of cluster merge in ECL & $2.2$ & $0.46$  & $0.02$ & $0.06$ \\
Detector resolution & 0.74 & 0.20 & 0.22 & 0.02 \\
Correction factor $R$  & $1.9$ &  $0.14$ & $0.04$ &  $0.04$ \\
Beam energy spread & negligible & negligible & negligible & negligible  \\ \hline 
Total & 7.0 & 1.1 & 0.76 & 0.24 \\ \hline \hline
\end{tabular}
\end{center}
\end{table}

\section{Results}

Based on the PDF in Eq.~\ref{generalpdfR}, we construct the NLL function, minimize it and obtain the Michel parameters $\bar{\eta}$ and $\xi\,\kappa$.
Since the sensitivity to $\bar{\eta}$ is suppressed by the factor of $m_l/m_\tau$,
 we extract it from the muon mode only.
 Using 832644 and 72768 selected events, respectively,
 for $\tau^+\tau^-\rightarrow (\pi^+\pi^0\bar{\nu})(e^-\nu\bar{\nu}\gamma)$ and
 $\tau^+\tau^-\rightarrow (\pi^+\pi^0\bar{\nu})(\mu^-\nu\bar{\nu}\gamma)$ candidates, we obtain 
\begin{align}
(\xi\kappa)^{(e)} &= -0.5 \pm 0.8 \pm 1.1, \\
\bar{\eta}^\mu &= -2.0 \pm 1.5 \pm 0.8,\\
(\xi\kappa)^{(\mu)} &= ~~0.8 \pm 0.5 \pm 0.2,
\end{align}
where the first error is statistical and the second
 is systematic. The results of $\xi\kappa$ are combined to give
\begin{equation}
\xi\kappa = 0.6 \pm 0.4 \pm 0.2.
\end{equation}

Figure~\ref{fitcont03} shows the contour the likelihood for $\tau\rightarrow \mu \nu \bar{\nu} \gamma $ events.
 As the shape suggests, a correlation between $\bar{\eta}$ and $\xi\kappa$ is small.
 The magnitude of the correlation coefficient determined by the error matrix is approximately $7\%$. 

\begin{figure}[]
{\centering {\includegraphics[width=8.0cm]{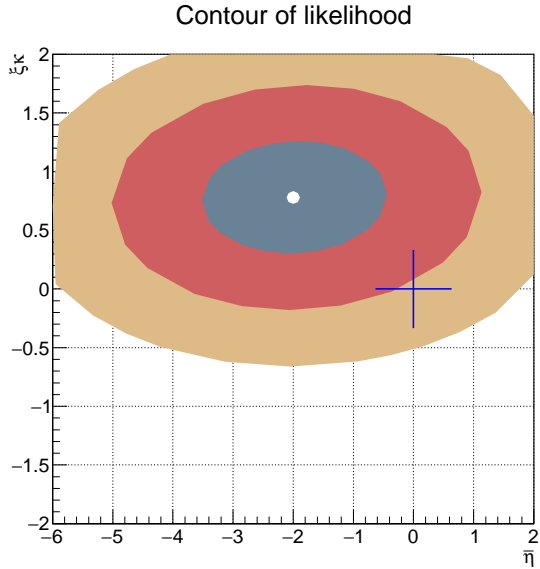}} }
\caption[]{\raggedright Contour of the likelihood for $\tau\rightarrow \mu \nu \bar{\nu} \gamma $ events.
 The circles indicate 1$\sigma$, 2$\sigma$ and 3$\sigma$ statistical deviations from inner to outer side, respectively.
 The cross is the SM prediction.
  }\label{fitcont03}
\end{figure}

\section{Summary}
We present preliminary results of a measurement of the Michel parameters $\bar{\eta}$ and $\xi\kappa$ of
 the $\tau$ using 703~f$\mathrm{b}^{-1}$ of data collected with the Belle detector at the KEKB $e^+e^-$ collider.
 These parameters are extracted from the radiative leptonic decay $\tau^-\rightarrow l^- \nu \bar{\nu} \gamma$
 and the tagging $\rho$ decay $\tau^+\rightarrow \rho^+ (\rightarrow \pi^+ \pi^0 )\bar{\nu}$ of the partner $\tau^+$ to exploit the spin-spin correlation in $e^+ e^- \to \tau^+ \tau^-$.
 Due to the poor sensitivity to $\bar{\eta}$ in the electron mode,
 this parameter is extracted only from $\tau^-\rightarrow \mu^- \nu \bar{\nu} \gamma$
 to give $\bar{\eta} = -2.0 \pm 1.5 \pm 0.8$.
 The product $\xi\kappa$ is measured using both decays $\tau^-\rightarrow l^- \nu \bar{\nu} \gamma$ ($l=e$ and $\mu$)
 to be $\xi\kappa = 0.6 \pm 0.4 \pm 0.2$. The first error is statistical and the second
 is systematic. This is the first measurement of both parameters for the $\tau$ lepton.
 These values are consistent with the SM expectation within the errors. 
 
 
\section{Acknowledgments}

We thank the KEKB group for the excellent operation of the
accelerator; the KEK cryogenics group for the efficient
operation of the solenoid; and the KEK computer group,
the National Institute of Informatics, and the 
PNNL/EMSL computing group for valuable computing
and SINET4 network support.  We acknowledge support from
the Ministry of Education, Culture, Sports, Science, and
Technology (MEXT) of Japan, the Japan Society for the 
Promotion of Science (JSPS), and the Tau-Lepton Physics 
Research Center of Nagoya University; 
the Australian Research Council;
Austrian Science Fund under Grant No.~P 22742-N16 and P 26794-N20;
the National Natural Science Foundation of China under Contracts 
No.~10575109, No.~10775142, No.~10875115, No.~11175187, No.~11475187
and No.~11575017;
the Chinese Academy of Science Center for Excellence in Particle Physics; 
the Ministry of Education, Youth and Sports of the Czech
Republic under Contract No.~LG14034;
the Carl Zeiss Foundation, the Deutsche Forschungsgemeinschaft, the
Excellence Cluster Universe, and the VolkswagenStiftung;
the Department of Science and Technology of India; 
the Istituto Nazionale di Fisica Nucleare of Italy; 
the WCU program of the Ministry of Education, National Research Foundation (NRF) 
of Korea Grants No.~2011-0029457,  No.~2012-0008143,  
No.~2012R1A1A2008330, No.~2013R1A1A3007772, No.~2014R1A2A2A01005286, 
No.~2014R1A2A2A01002734, \\ No.~2015R1A2A2A01003280 , No. 2015H1A2A1033649;
the Basic Research Lab program under NRF Grant No.~KRF-2011-0020333,
Center for Korean J-PARC Users, No.~NRF-2013K1A3A7A06056592; 
the Brain Korea 21-Plus program and Radiation Science Research Institute;
the Polish Ministry of Science and Higher Education and 
the National Science Center;
the Ministry of Education and Science of the Russian Federation and
the Russian Foundation for Basic Research, grant 15-02-05674;
the Slovenian Research Agency;
Ikerbasque, Basque Foundation for Science and
the Euskal Herriko Unibertsitatea (UPV/EHU) under program UFI 11/55 (Spain);
the Swiss National Science Foundation; 
the Ministry of Education and the Ministry of Science and Technology of Taiwan;
and the U.S.\ Department of Energy and the National Science Foundation.
This work is supported by a Grant-in-Aid from MEXT for 
Science Research in a Priority Area (``New Development of 
Flavor Physics'') and from JSPS for Creative Scientific 
Research (``Evolution of Tau-lepton Physics'').

\appendix

\section{Appendix~A: Differential decay width of $\tau \rightarrow l \nu \bar{\nu} \gamma$ \label{app_dif_sig}}

The general differential cross section of $\tau \rightarrow l \nu \bar{\nu} \gamma$ decay is expressed
 as a sum of the two terms:

\begin{equation}
\frac{\mydif\Gamma( \tau^\mp \rightarrow l^\mp \nu_{\tau} \bar{\nu_{l}} \gamma )}{\mydif E^*_l \mydif \Omega^*_l \mydif E^*_\gamma \mydif \Omega^*_\gamma } 
=  A \mp  \bvec{B}\cdot \bvec{S}_{\tau^\mp},
\end{equation}
where $A$ and $\bvec{B}$ represent spin-independent and spin-dependent terms. These terms are functions of dimensionless kinematic parameters $x,y$ and $d$ defined as:
\begin{align}
r &= \frac{m_l}{m_{\tau}}, \\
x &= \frac{2E_{l}^*}{m_{\tau}},~~(2r < x < 1+r^2) \\ 
y &= \frac{2E_{\gamma}^*}{m_{\tau}},~~(0 < y < 1-r) \\ 
d &= 1-\beta_{l}^* \cos \theta_{l \gamma}^*, \\ 
y &< \frac{2(1+r^2-x)}{2-x+\cos \theta_{l\gamma}^* \sqrt{x^2-4r^2}}. \label{Dynamicrelation} 
\end{align}
$A$ and $\bvec{B}$ are parametrized by the Michel parameters $\rho$, $\eta$, $\xi$, $\xi\delta$, $\bar{\eta}$, $\eta^{\prime \prime}$ and $\xi\kappa$. 

\begin{align}
A(x,y,d)&= \frac{4\alpha G_F^2 m_{\tau}^3 }{3(4\pi)^6 } \cdot \beta_l \sum_{i=0,1\ldots 5}{F_{i}r^i}, \\
\bvec{B}(x,y,d)&= \frac{4 \alpha G_F^2 m_{\tau}^3 }{3(4\pi)^6 } \cdot \beta_l \sum_{i=0,1\ldots 5}{(\beta_{l}^*G_{i}\bvec{n}_{l}^{*}+H_{i}\bvec{n}_{\gamma}^{*})r^i} \label{sigeq_1},
\end{align}
where $\bvec{n}_l^*$ and $\bvec{n}_\gamma^*$ are normalized directions of lepton and photon in the $\tau$ rest frame, respectively.
The $F_i$, $G_i$ and $H_i$ ($i=0,1\ldots 5$) are functions of $x$, $y$, $d$ and $r$ and their explicit formulae are given in the Appendix of Ref.~\cite{michel_form_arvzov}.
\section{Appendix~B: Differential decay width of $\tau \rightarrow \rho \nu$}

We use the CLEO model to define the differential decay width of $\tau^\pm \rightarrow \rho^\pm \nu$ decay.
 This is expressed as a sum of the spin-independent and spin-dependent parts~\cite{cite_CLEO_rho,citeTAUOLArho}:
\begin{align}
&\hspace{3cm}\frac{{\mathrm{d}\Gamma(\tau^\pm\rightarrow}
  \pi^\pm\pi^0\nu)}{\mydif \Omega_{\rho}^*\mydif m^2 \mydif \widetilde{\Omega}_{\pi}} = A^\prime \mp \bvec{B}^\prime \cdot \bvec{S_{\tau^\pm}}, \\
 A^\prime =& \frac{G_F^2 |V_{ud}|^2}{(4\pi)^5 } \cdot \Big[2(E_{\pi}^*-E_{\pi^0}^*)(p_{\nu}\cdot q)-E_{\nu}^*q^2\Big] \cdot \mathrm{BPS},\\
\bvec{B}^\prime =& \frac{G_F^2 |V_{ud}|^2}{(4\pi)^5 } \cdot  \Big[
  \bvec{ P}_{\pi}^* \left\{ ( q\cdot q) +2(p_{\nu}\cdot q)\right\}
+\bvec{P}_{\pi^0}^* \left\{ ( q\cdot q) -2(p_{\nu}\cdot q)\right\} 
 \Big] \cdot \mathrm{BPS}, \label{sigeq_2}
\end{align}
where $V_{ud}$ is the corresponding element of the Cabibbo-Kobayashi-Maskawa matrix and $q$ is a four-vector defined by $q=p_{\pi}-p_{\pi^0}$.
 The factor BPS stands for a square of a relativistic Breit-Wigner function and a Lorentz-invariant phase space and it is calculated from the following formulae.
\begin{align}
\mathrm{BPS} =& \left|\mathrm{BW}(m^2) \right|^2\ \left( \frac{2P_{\rho}^{*}(m^2)}{m_\tau} \right) \left(\frac{2\tilde{P}_{\pi}(m^2)}{m_{\rho}} \right), ~~~
\mathrm{BW}(m^2) =\frac{\mathrm{BW}_\rho+\beta \mathrm{BW}_{\rho^{\prime}}}{1+\beta}, \\
\mathrm{BW}_{\rho}\left(m^2\right) =& \frac{m_{\rho 0}^{2}}{m_{\rho 0}^{2}-m^{2}-im_{\rho 0}\Gamma_{\rho }\left(m^{2}\right)}
,~~~ \Gamma_{\rho}\left(m^2\right) =\displaystyle \Gamma_{\rho 0}\frac{m_{\rho 0}}{\sqrt{m^{2}}} \left(\frac{\tilde{P_{\pi}}\left(m^{2}\right)}{\tilde{P_{\pi}}\left(m_{\rho 0}^{2}\right)} \right)^{3}, \\
\mathrm{BW}_{\rho^{\prime}}\left(m^2\right) =& \frac{m_{\rho^{\prime}0}^{2}}{m_{\rho^{\prime} 0}^{2}-m^{2}-im_{\rho^{\prime}0}\Gamma_{\rho^{\prime}}\left(m^{2}\right)}
,~~~\Gamma_{\rho^{\prime } }\left(m^2\right)=\displaystyle \Gamma_{\rho^{\prime } 0}\frac{m_{\rho^{\prime }0}}{\sqrt{m^{2}}}\left(\frac{\tilde{P_{\pi}}\left(m^{2}\right)}{\tilde{P_{\pi}}\left(m_{\rho^{\prime } 0}^{2}\right)}\right)^{3}, \\
\hspace{2cm}&\tilde{P}_{\rho}^*(m^2) = \frac{m_\tau^2- m^2}{2 m_\tau} \\
\hspace{2cm}&\tilde{P}_{\pi}(m^2) = \frac{\sqrt{\left[m^2-(m_\pi-m_{\pi^0})^2\right]\left[m^2-(m_\pi +m_{\pi^0})^2\right]}}{2m}.
\end{align}
A factor $\mathrm{BW}_a$ ($a = \rho$ or $\rho^\prime$) represents the Breit-Wigner function associated with a $\rho$ or a $\rho^\prime$ resonance mass shape and the parameter $\beta$ specifies their relative coupling.
 $m_{\rho0}$ and $m_{\rho^{\prime}0}$ are nominal masses of two resonance states and $\Gamma_{\rho0}$ and $\Gamma_{\rho^{\prime}0}$ are their nominal total decay widths.


\end{document}